\documentclass[iop,apj,numberedappendix,appendixfloats,twocolappendix]{emulateapj}

\usepackage{amsmath,amssymb}
\usepackage{color}

\newcommand{\bfB}{\mbox{\boldmath$B$}}

\newcommand\calP{{\cal P}}
\newcommand\calA{{\cal A}}

\newcommand{\Bbar}{{\bar B}}

\newcommand{\rhobar}{{\bar \rho}}
\newcommand{\bfb}{\mbox{\boldmath$b$}}

\newcommand{\bfe}{\mbox{\boldmath$e$}}
\newcommand{\bfu}{\mbox{\boldmath$u$}}

\newcommand{\invHrho}{H_\rho^{-1}}
\newcommand{\invHB}{H_B^{-1}}
\newcommand{\drm}{\mathrm{d}}
\newcommand{\erm}{\mathrm{e}}
\newcommand{\Orm}{\mathrm{O}}
\newcommand{\orm}{\mathrm{o}}
\newcommand{\maxrm}{\mathrm{max}}
\shorttitle{Short wavelength magnetic buoyancy instability}
\shortauthors{Mizerski, Davies \& Hughes}

\begin{document}

\title{Short wavelength magnetic buoyancy instability}

\author{K. A. Mizerski$^{1,2}$, C. R. Davies$^1$ and D. W. Hughes$^1$}
\affil{$^{1}$Department of Applied Mathematics, University of Leeds, Leeds LS2 9JT, U.K.\\
$^2$Department of Magnetism, Institute of Geophysics, Polish Academy of Sciences, ul. Ksiecia Janusza 64, 01-452 Warsaw, Poland}
\email{kamiz@igf.edu.pl, tina@maths.leeds.ac.uk, d.w.hughes@leeds.ac.uk}

\begin{abstract}
Magnetic buoyancy instability plays an important role in the evolution of astrophysical magnetic fields. Here we revisit the problem introduced by \citet{Gilman_1970} of the short wavelength linear stability of a plane layer of compressible isothermal fluid permeated by a horizontal magnetic field of strength decreasing with height. Dissipation of momentum and magnetic field is neglected. By the use of a Rayleigh-Schr\"odinger perturbation analysis, we explain in detail the limit in which the transverse horizontal wavenumber of the perturbation, denoted by $k$, is large (i.e.\ short horizontal wavelength) and show that the fastest growing perturbations become localized in the vertical direction as $k$ is increased. The growth rates are determined by a function of the vertical coordinate $z$ since, in the large $k$ limit, the eigenmodes are strongly localized in the vertical direction. We consider in detail the case of two-dimensional perturbations varying in the directions perpendicular to the magnetic field, which, for sufficiently strong field gradients, are the most unstable. The results of our analysis are backed up by comparison with a series of initial value problems. Finally we extend the analysis to three-dimensional perturbations. 
\end{abstract}


\section{Introduction}
\label{sec:intro}
Instabilities due to magnetic buoyancy are driven by stratified horizontal magnetic fields in compressible plasmas \citep{Newcomb_1961}, and may thus occur in a range of astrophysical settings --- for example, in stars, in accretion disks and in the interstellar medium \citep[see][]{Parker_1979, Choudhuri_1998}. Of particular recent interest is the idea that such instabilities are responsible for the break-up and escape of predominantly toroidal magnetic field from the solar tachocline \citep[see][]{Hughes_2007}, producing the stitches of field that eventually appear at the surface as active regions.

One of the earliest studies of magnetic buoyancy instability was by \cite{Gilman_1970}, who considered the linear instability of a magnetohydrostatic atmosphere with a vertically stratified, horizontal magnetic field. Motivated by the extreme parameter values that pertain astrophysically, Gilman considered the case of an inviscid, perfectly conducting gas in which the thermal relaxation is sufficiently rapid that the temperature can be specified for all times. He anticipated that, under these assumptions, the fastest growing perturbations would be infinitesimally narrow in the horizontal direction perpendicular to the imposed horizontal magnetic field. Interestingly, this approach leads to a `dispersion relation' relating the (possibly complex) frequency to the horizontal wavenumber, but one in which the coefficients are dependent on the vertical coordinate $z$. Thus one may formally associate a different dispersion relation with each height. Mathematically, however, the problem can be posed as a two-point boundary value problem with well-defined (constant) eigenvalues. The aim of this paper is to clarify the connection between these two ostensibly rather different approaches.

The analysis, although new to this problem and, as far as we are aware, to the study of magnetohydrodynamic instabilities, may be regarded as essentially a Rayleigh-Schr\"odinger perturbation analysis exploiting the large transverse horizontal wavenumber. In geophysical fluid dynamics, inertial instabilities of a rotating, stratified flow with arbitrary horizontal cross-stream shear have been analyzed using this technique by \cite{Griffiths_2008}, who considered perturbations that are highly localized in the horizontal cross-stream direction. The alternative to this boundary layer (or internal layer) type approach is a WKB analysis, as used in magnetohydrodynamics by \citet{TP_1996} and \citet{Ogilvie_1998}, who considered highly localized (in radius) magnetic instabilities in accretion disks. Each method has its strengths: on the one hand, WKB analysis is more general; on the other, the boundary layer approach is mathematically simpler and more physically appealing.

The paper is organized as follows. The governing equations describing the magnetic buoyancy instability of a layer of gas with a vertically stratified horizontal magnetic field are set out in Section~\ref{sec:Math_form}, together with the formulation of the problem both as a two-point boundary value problem and as one yielding a `depth-dependent dispersion relation' with growth rate $\sigma(z)$. For simplicity we first restrict attention to two-dimensional (interchange) perturbations, for which the magnetic field remains unidirectional. In Section~\ref{sec:example}, in order to elucidate the key aspects of the analysis, we revisit the problem of the quantum harmonic oscillator, which shares an important common feature with the problem of magnetic buoyancy instability at high wavenumber. In Section~\ref{sec:MBE}, we apply a high wavenumber asymptotic analysis to the governing MHD equation, considering in detail two distinct cases, depending on whether $\sigma(z)$ is maximized strictly within the layer of gas (where $\sigma$ is locally quadratic in $z$) or whether the maximum occurs at the boundary (with $\sigma$ locally linear in $z$). As a result of this analysis we are able to reconcile the idea of a depth-dependent dispersion relation with the solutions of the two-point boundary eigenvalue problem. In Section~\ref{sec:TE} we consider the instability from the different standpoint of an initial value problem, calculating how an initial perturbation evolves with time, and demonstrating the relation of this solution to those derived in Section~\ref{sec:MBE}. In Section~\ref{sec:3D} we show how the high wavenumber analysis carries over to the case of fully three-dimensional perturbations. The concluding discussion is contained in Section~\ref{sec:Disc}.

\section{Mathematical formulation}
\label{sec:Math_form}

Following \cite{Gilman_1970}, we consider magnetic buoyancy instabilities under the assumptions that the gas is isothermal, inviscid and electrically perfectly conducting. The main argument of the paper can be advanced even when the system is rather simple, so here we choose to ignore the effects of rotation. By adopting the layer depth $d$, the free fall time $\sqrt{d/g}$ and the free fall velocity $\sqrt{gd}$ as units of length, time and velocity respectively (where $g$ is the acceleration of gravity, assumed constant), the equations of motion, induction and mass conservation, together with the perfect gas law, take the following dimensionless form:
\begin{equation}
\rho \left( \frac{\partial\bfu}{\partial t} + \left(\bfu\cdot\nabla\right) \bfu \right) = -\mathcal{P}\nabla p - \rho\hat{\mathbf{e}}_{z} + \Lambda\left(\nabla\times\bfB\right)\times\bfB ,
\label{eq:NS}
\end{equation}
\begin{equation}
\frac{\partial\bfB}{\partial t} + \left(\bfu\cdot\nabla\right)\bfB = \left(\bfB\cdot\nabla\right)\bfu - \bfB \left(\nabla\cdot\bfu\right) ,
\label{eq:IND}
\end{equation}
\begin{equation}
\nabla \cdot \bfB = 0 ,
\label{eq:divB}
\end{equation}
\begin{equation}
\frac{\partial\rho}{\partial t} + \nabla\cdot\left(\rho\bfu\right)=0 \, ,
\label{eq:MASS}
\end{equation}
\begin{equation}
p=\alpha\rho,
\label{eq:EQSTATE}
\end{equation}
where
\begin{equation}
\mathcal{P} = \frac{p_s}{\rho_s gd}, \quad \alpha = \frac{\rho_s RT_0}{p_s},
\quad
\Lambda=\frac{B_s^2}{\mu_0\rho_s gd}=\beta^{-1} \mathcal{P}.
\label{eq:param_def}
\end{equation}
Here $T_0$ is the constant temperature of the system, $R$ is the gas constant, $\mu_0$ is the permeability of free space, $p_{s}$, $\rho_{s}$ and $B_{s}$ are representative values of the pressure, density and magnetic field respectively. The plasma $\beta$, here defined by
\begin{equation}
\beta = \frac{\mu_0 p_s}{B_s^2} ,
\label{eq:plasma_beta}
\end{equation}
represents the ratio of the gas pressure to twice the magnetic pressure. Additionally, we define the non-dimensional isothermal speed of sound $\mathcal{U}_S=\sqrt{RT_0/gd}$, the non-dimensional Alfv\'en speed $\mathcal{U}_A=B_s/\sqrt{\mu_0\rho_s gd}$ and $\mathcal{U}_p=\sqrt{p_s/\rho_s gd}$. The following relations then hold between the various parameters:
\begin{equation}
\mathcal{P}=\mathcal{U}_{p}^2,\qquad\Lambda=\mathcal{U}_{A}^2,\qquad\alpha=\frac{\mathcal{U}_{S}^2}{\mathcal{U}_{p}^2}.\label{eq:Param_relations}\end{equation}
We now consider a stationary equilibrium layer of gas in the (dimensionless) region $0 \le z \le 1$ with a horizontal, depth-dependent magnetic field,
\begin{equation}
\bfB = \Bbar \left( z \right) \bfe_x.
\label{eq:BS_field}
\end{equation}
The pressure and density of the static basic state are determined by the two coupled equations
\begin{equation}
\mathcal{P} \frac{\drm \bar{p}}{\drm z} = -\frac{\Lambda}{2}  \frac{\drm \Bbar^2}{\drm z} - \rhobar  ,
\label{eq:isotherm_BS_eq_1}
\end{equation}
\begin{equation}
\bar{p} = \alpha \rhobar ,
\label{eq:isotherm_BS_eq_2}
\end{equation}
from which the density distribution is given by,
\begin{align}
\nonumber
\rhobar(z) = &\rhobar(0) \erm^{-z/\calP \alpha} -
\frac{\Lambda}{2 \calP \alpha} \left( \Bbar^2(z) - \Bbar^2(0) \erm^{-z/\calP \alpha}  \right) \\
& + \frac{\Lambda}{2 \calP^2 \alpha^2}  \erm^{-z/\calP \alpha} \int_0^z \erm^{-z/\calP \alpha} \Bbar^2(z) \drm z .
\label{eq:BS_rho}
\end{align}

In order to retain as much simplicity as possible, we shall first assume that the perturbations to the basic state are two-dimensional, varying in the directions perpendicular to the equilibrium magnetic field (\textit{interchange modes}) (three-dimensional perturbations are considered in Section~\ref{sec:3D}). Thus we adopt perturbations of the following form:
\begin{subequations}
\begin{equation}
\bfu = (0, v(z), w(z)) \erm^{\sigma t+ i ky} ,
\label{eq:pert_form_intro_u}
\end{equation}
\begin{equation}
\bfb = (b_x(z), 0, 0)\erm^{\sigma t+ i ky},
\label{eq:pert_form_intro_b}
\end{equation}
\begin{equation}
p = \tilde{p} (z) \, \erm^{\sigma t+i ky}\,, \qquad \rho = \tilde{\rho} (z) \, \erm^{\sigma t+ i ky}.
\label{eq:pert_form_intro_p_rho}
\end{equation}
\end{subequations}
Introducing these into equations~(\ref{eq:NS}) -- (\ref{eq:EQSTATE}) (equation~(\ref{eq:divB}) is trivially satisfied), with the basic state given by expressions~(\ref{eq:BS_field}), (\ref{eq:isotherm_BS_eq_2}) and (\ref{eq:BS_rho}), and then linearizing, leads to the following set of equations:
\begin{equation}
\sigma\rhobar v=-i k\left(\mathcal{P}\tilde{p} + \Lambda \Bbar b_x \right) ,
\label{eq:pert_v}
\end{equation}
\begin{equation}
\sigma\rhobar  w = -\frac{\drm}{\drm z} \left( \mathcal{P} \tilde{p} + \Lambda \Bbar b_x \right) -\tilde{\rho},
\label{eq:pert_w}
\end{equation}
\begin{equation}
\sigma b_x=-i k \Bbar v - \frac{\drm}{\drm z}\left( \Bbar w \right),
\label{eq:pert_b}
\end{equation}
\begin{equation}
\sigma \tilde{\rho} = - i k \rhobar  v - \frac{\drm}{\drm z} \left( \rhobar  w \right),
\label{eq:pert_rho}
\end{equation}
\begin{equation}
\tilde{p}=\alpha\tilde{\rho} .
\label{eq:pert_p}
\end{equation}
Equations~(\ref{eq:pert_v}) -- (\ref{eq:pert_p}) can be manipulated to give the following second-order ordinary differential equation for the vertical velocity $w$:
\begin{align}
\sigma^2 \rhobar w &= \frac{k^2 \Lambda \Bbar^2}{\sigma^2 + k^2 F(z)} \left( \frac{1}{H_\rho} - \frac{1}{H_B} \right) w \nonumber \\
&+ \frac{\sigma^2 \rhobar}{\sigma^2 + k^2 F(z)} \left( w \frac{1}{H_\rho} + \frac{\drm w}{\drm z} \right) + \nonumber \\
&+ \frac{\drm}{\drm z} \left( \frac{\sigma^2 \rhobar F(z)}{\sigma^2 + k^2 F(z)} \frac{\drm w}{\drm z} \right. \nonumber \\
&\left. - \frac{\sigma^2}{\sigma^2 + k^2 F(z)} \left( \Lambda \Bbar^2 \left( \frac{1}{H_\rho} - \frac{1}{H_B} \right) -\rhobar F(z) \frac{1}{H_\rho} \right) w \right) ,
\label{eq:bvp}
\end{align}
where the ($z$-dependent) inverse scale heights for density and magnetic field are given by $\invHrho (z) = \rhobar ^{-1} \drm \rhobar / \drm z$, $\invHB (z) = \Bbar^{-1} \drm \Bbar / \drm z$, and where $F(z) = \alpha\mathcal{P} + \Lambda \Bbar^2 / \rhobar$. Thus the MHD stability problem reduces to a two-point boundary problem, with boundary conditions on the vertical velocity at the horizontal boundaries; the simplest choice is to impose impermeability, namely $w=0$ at $z=0, 1$. The growth rate $\sigma$ is determined as an eigenvalue of the problem. For a general $z$-dependent basic state, equation~(\ref{eq:bvp}) requires a numerical solution.

Gilman's (1970) analysis proceeded by considering the limit as $k \to \infty$ in the governing equations (here (\ref{eq:pert_v}) -- (\ref{eq:pert_p})), but with no explicit assumption about the locality in $z$ of the perturbations. This leads, from (\ref{eq:pert_v}), to the vanishing of the total pressure perturbation, thus giving a direct relation between $b_x$ and $\tilde p$. Similarly, from equation~(\ref{eq:pert_w}), the vanishing of the total pressure perturbation leads to a relation between $w$ and $\tilde \rho$. Thus, using also equation~(\ref{eq:pert_p}), $b_x$, $\tilde \rho$ and $\tilde p$ can all be expressed in terms of $w$. On eliminating the finite product $kv$ between equations~(\ref{eq:pert_b}) and (\ref{eq:pert_rho}), the terms in $\drm w/\drm z$ disappear. Substitution for $b_x$, $\tilde \rho$ and $\tilde p$ in terms of $w$ then leads to the following expression:
\begin{equation}
\left[ \sigma^2 F(z) + \frac{\Lambda \Bbar^2}{\rhobar} \left( \frac{1}{H_B} - \frac{1}{H_\rho} \right) \right] w = 0 .
\label{eq:Gilman}
\end{equation}
This is a simplified version of expression~(14) in \cite{Gilman_1970}, excluding the effects of rotation and three-dimensional perturbations. Expression~(\ref{eq:Gilman}) may be regarded as a depth-dependent dispersion relation, in that it implies that the instability growth rate $\sigma$ (a constant) takes a different value at each height $z$. With this interpretation, it leads to the following criterion for instability ($\sigma^2 > 0$):
\begin{equation}
- \frac{\drm}{\drm z} \ln \left( \frac{\Bbar}{\rho} \right) > 0.
\label{eq:inst_criterion}
\end{equation}
This is consistent with the result of \cite{Acheson_1979}, obtained under the assumption that perturbations are localized in both the horizontal \textit{and} vertical directions; in the derivation of expression~(\ref{eq:Gilman}), however, there is no explicit assumption of locality in $z$. Expression~(\ref{eq:Gilman}) is clearly consistent with formally letting $k \to \infty$ in~(\ref{eq:bvp}); our aim in this paper is to relate the solutions of the full governing ordinary differential equation~(\ref{eq:bvp}) to expressions~(\ref{eq:Gilman})  and (\ref{eq:inst_criterion}). For later work it will be helpful formally to define the function $\sigma(z)$ by
\begin{equation}
\sigma(z)=  \left( \frac{\Lambda \Bbar^2}{\rhobar F(z)} \left( \frac{1}{H_\rho} - \frac{1}{H_B} \right) \right)^{1/2} .
\label{eq:sigma(z)}
\end{equation}
We shall retain the symbol $\sigma$ to denote the true eigenvalues.

\section{An example problem} \label{sec:example}

In this section we introduce a simplified example problem designed to mimic some of the key properties of the perturbation equations for the magnetic buoyancy instability in the short wavelength limit. We consider a linear PDE, first-order in time and second-order in space ($z$), containing a parameter $k$ to represent the wavenumber of the system. With the introduction of a growth rate $\sigma$ this becomes a second-order ODE analogous to (\ref{eq:bvp}). Just as for the full ODE~(\ref{eq:bvp}), our example reduces to a purely algebraic equation determining $\sigma$ as a function of $z$ in the formal limit of $k\to\infty$ (and where the gradient terms are small in comparison with this limit). The example highlights the meaning of the growth rate in such a limit.

We consider the following PDE for the function $f(z,t)$ on the spatial domain $0 \le z \le 1$:
\begin{equation}
\frac{\partial f}{\partial t} = \left[ \sigma_{\maxrm} - (z-z_{\maxrm})^2 \right] f + \frac{1}{k^2} \frac{\partial ^2 f}{\partial z^2} .
\label{eq:exPDE}
\end{equation}
For simplicity, we impose the  boundary conditions $f(0,t) = f(1,t) =0$, although this specific choice is not crucial for the arguments that follow. Equation~(\ref{eq:exPDE}) contains the parameters $\sigma_{\maxrm}$, which we shall assume to be positive, and $z_{\maxrm}$, which we assume satisfies $0 < z_{\maxrm} < 1$.

On expressing the time dependence of the function $f$ as $f\propto e^{\sigma t}$, equation~(\ref{eq:exPDE}) becomes the second-order ODE
\begin{equation}
\frac{\drm^2 f}{\drm z^2} - k^2 \left[ (\sigma - \sigma_{\maxrm}) + (z - z_{\maxrm})^2 \right] f = 0,
\label{eq:exODE}
\end{equation}
with boundary conditions $f(0)=f(1)=0$. As $k$ is increased, the derivative term in equation~(\ref{eq:exODE}) becomes increasingly unimportant, except in the regions where the coefficient of $f$ could be comparable with the second derivative. In the large $k$ limit described earlier, equation~(\ref{eq:exODE}) reduces to
\begin{equation}
\sigma = \sigma_{\maxrm} - (z-z_{\maxrm})^2,
\label{eq:exGR}
\end{equation}
an equation for the growth rate as a function of $z$. This quadratic function is clearly maximized at $z=z_{\maxrm}$, with $\sigma=\sigma_{\maxrm}$. We might therefore expect to see the fastest growing solutions of equation~(\ref{eq:exODE}) become increasingly localized about $z=z_{\maxrm}$ as $k$ is increased. For the particular equation~(\ref{eq:exODE}) this idea can be put on a rigorous footing.

Equation~(\ref{eq:exODE}) can be recast in the standard form of the parabolic cylinder equation \begin{equation}
\frac{\drm^2 f}{\drm x^2} - \left( \frac{1}{4} x^2 + a \right) f = 0
\label{eq:exPC}
\end{equation}
by writing $x=(z-z_{\maxrm})\sqrt{2k}$ and $a=k(\sigma-\sigma_{\maxrm})/2$. The boundary conditions are now $f=0$ at both $x_1 = -z_{\maxrm}\sqrt{2k}$ and $x_2 = (1-z_{\maxrm})\sqrt{2k}$. 

\begin{figure}
\plotone{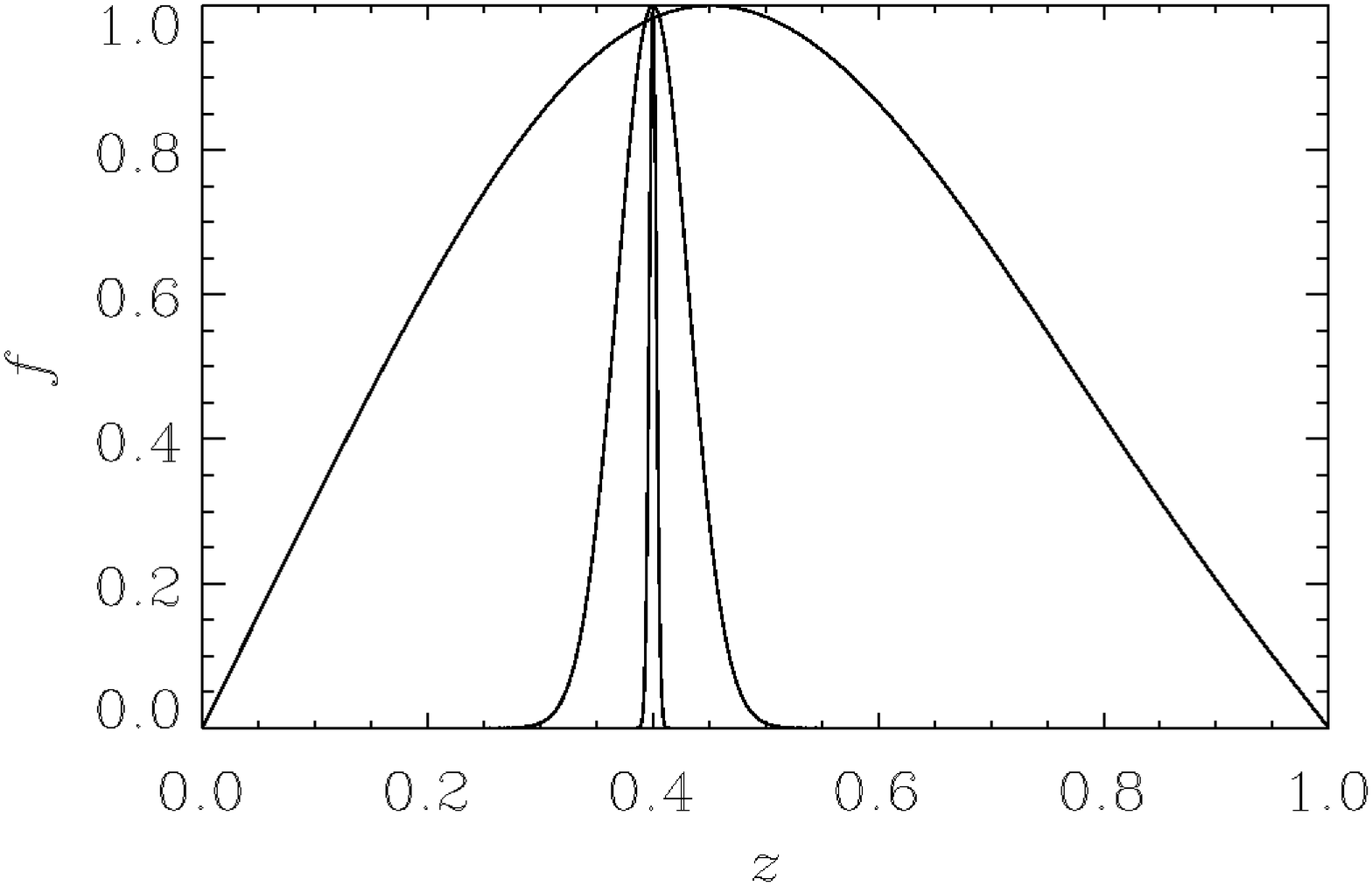}
\\
\plotone{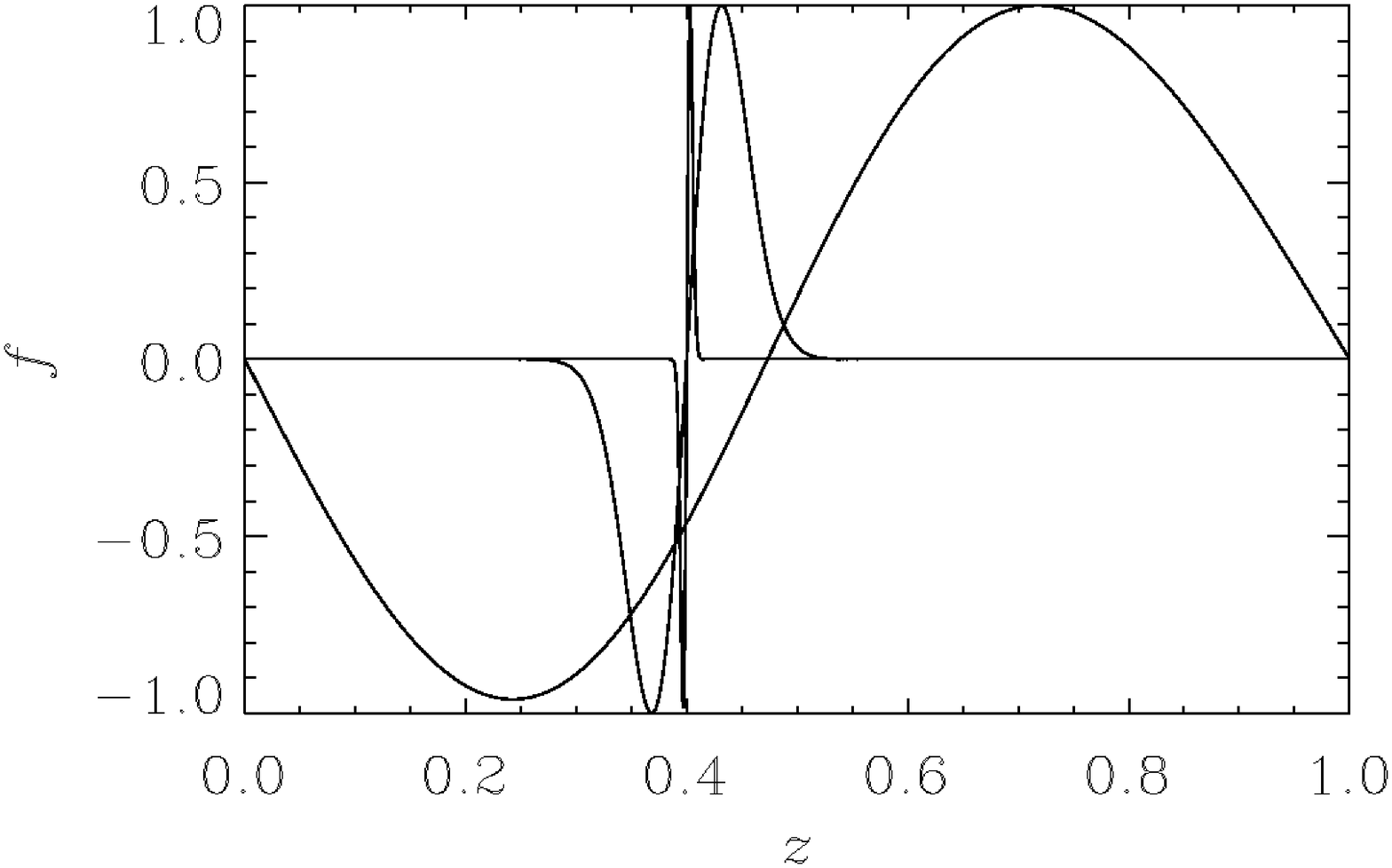}
\caption{Eigenfunctions of equation~(\ref{eq:exPC}) for (a) $a=-1/2$, (b) $a=-3/2$, for $k= 10$, $10^3$ and $10^5$. The eigenfunctions become increasingly peaked about $z= z_{\maxrm}=0.4$ as $k$ is increased.}
\label{fig:f_eigenmode}
\end{figure}

To determine the permissible values of $a$ we note that $|x_1|, |x_2| \gg 1$ in the large $k$ limit, provided that $z_{\maxrm}\gg k^{-1/2}$ and $1-z_{\maxrm}\gg k^{-1/2}$. The problem is then the familiar one of a quantum harmonic oscillator for the wave function $\psi$ with $\psi \to 0$ as $x \to \pm \infty$. The most general solution to equation~(\ref{eq:exPC}) \citep[see, for example,][]{BenderOrszag} may be expressed in terms of parabolic cylinder functions $D_\nu$ as
\begin{equation}
f(x)=c_1 D_\nu (x) + c_2 D_{-\nu -1}(-ix) ,
\label{eq:qho}
\end{equation}
where $\nu = -1/2-a$. For $|\arg z| < 3 \pi/4$, the asymptotic form of $D_\nu(z)$ as $z \to \infty$ is given by
\begin{align}
\nonumber
D_\nu (z) \sim z^\nu e^{-z^2/4}
&\left( 1 - \frac{\nu(\nu-1)}{2z^2} + \right. \\
&\left. \frac{\nu(\nu-1)(\nu-2)(\nu-3)}{2\times 4 z^4} - \cdots \right) .
\label{eq:D_asymp_1}
\end{align}
Thus $D_{-\nu -1}(-ix)$ grows exponentially as $x \to \infty$ for all $\nu$; hence to satisfy $f \to 0$ as $x \to \infty$ it follows that $c_2=0$. Since $D_\nu (x) \to 0$ as $x \to \infty$, the most general solution of equation~(\ref{eq:qho}) satisfying $f \to 0$ as $x \to \infty$ is $f = c_1 D_\nu(x)$. Now for $3\pi/4 < \arg z < 5 \pi/4$, the asymptotic form of $D_\nu(z)$ as $z \to \infty$ becomes
\begin{align}
\nonumber
D_\nu (z) \sim &- \frac{(2 \pi)^{1/2}}{\Gamma(-\nu)} e^{i \pi \nu} z^{-\nu-1} e^{z^2/4}
\left( 1 + \frac{(\nu+1)(\nu+2)}{2z^2} + \right. \\
& \left. \frac{(\nu+1)(\nu+2)(\nu+3)(\nu+4)}{2 \times 4z^4} + \cdots \right) .
\label{eq:D_asymp_2}
\end{align}
Hence $D_\nu(x)$ grows exponentially as $x \to - \infty$ unless $\Gamma(-\nu)$ is infinite; this occurs only if $\nu$ is a non-negative integer. Therefore the only allowable values of $a$ are $a=-1/2$, $-3/2$, $-5/2, \, \ldots$. When $\nu = n$ $= 0, 1, 2, \ldots$,
\begin{equation}
D_n(x)= \exp(-x^2/4) \textrm{He}_n (x) ,
\label{eq:Dn}
\end{equation}
where $\textrm{He}_n (x)$ is the $n$th-degree Hermite polynomial. The first few $\textrm{He}_n (x)$ are given by $\textrm{He}_0 (x) =1$,  $\textrm{He}_1 (x) =x$, $\textrm{He}_2 (x) =x^2-1$, $\textrm{He}_3 (x) =x^3-3x$; in general $\textrm{He}_n (x)$ is even (odd) in $x$ for even (odd) $n$. Thus there is a solution for each $n\in\mathbb{N}\cup\{0\}$, with $n+1$ turning points. As $k$ is increased, the spacing between the turning points decreases and the functions become increasingly localized about $z=z_{\maxrm}$. This is illustrated by Figure~\ref{fig:f_eigenmode}, which shows the eigenfunctions for $n=0$ and $n=1$ for increasing values of $k$.

\begin{figure}
\plotone{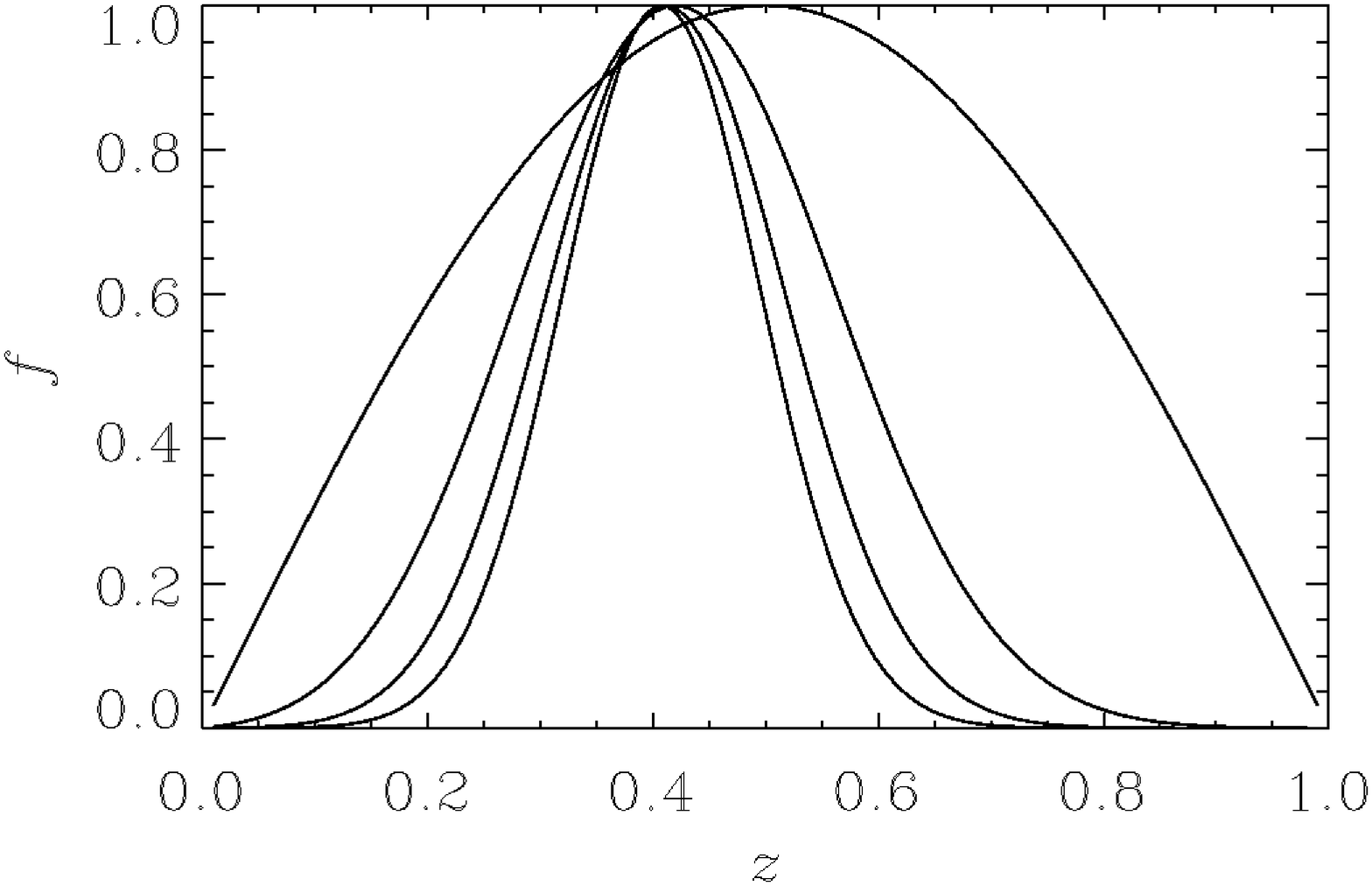}
\\
\plotone{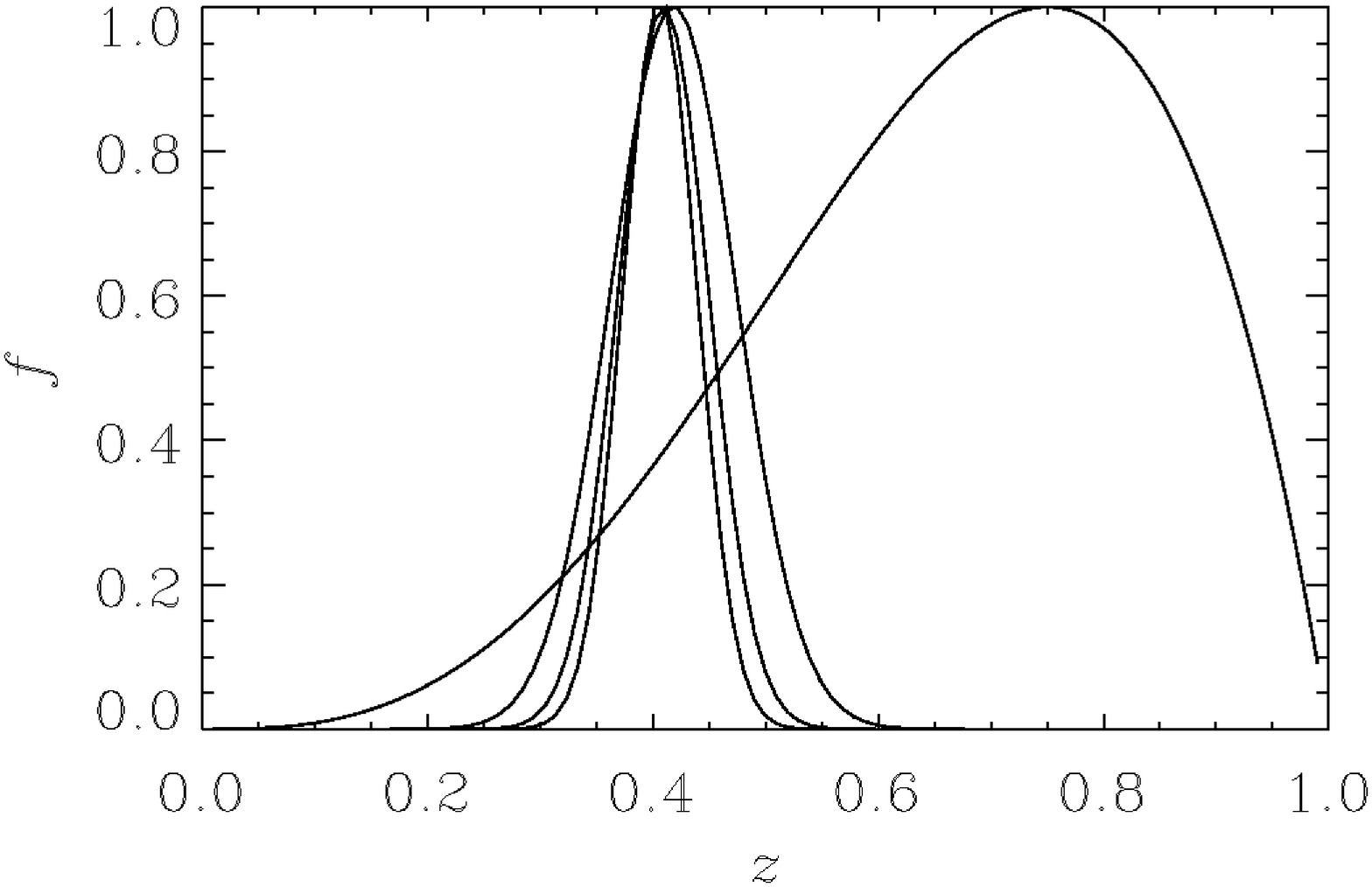}
\caption{Temporal evolution of the solution of equation~(\ref{eq:exPDE}) towards the maximally growing $a=-1/2$ eigenfunction for a moderately high wavenumber, $k=10^5$. In (a) the initial condition is $f(z)= \sin \pi z$ and the solution is shown at $t=0$, $20$, $40$ and $60$. In (b) the initial condition is $f(z)= z(1-z)^3$ and the solution is shown at $t=0$, $150$, $300$ and $450$. All solutions are normalized so that $\max(f)=1$.}
\label{fig:f_ivp}
\end{figure}

The eigenvalue analysis above can now be related to the original evolution PDE (\ref{eq:exPDE}). Appealing to the completeness of the eigenfunctions of the linear operator, the solution $f(z,t)$ can be expressed as a sum of exponentially growing eigenfunctions (\ref{eq:Dn}). Thus, from an arbitrary initial condition $f(z,0)$, the long term solution will have the spatial form $D_0(x)$, corresponding to the maximum growth rate of $\sigma = \sigma_{\maxrm} - 1/k$. The eigenfunction is peaked at $x=0$, i.e.\ at $z=z_\maxrm$. This is illustrated by Figure~\ref{fig:f_ivp}, which shows the temporal evolution of the spatial dependence of $f(z,t)$ starting from two initial conditions.

\section{The Magnetic Buoyancy Equation} \label{sec:MBE}

Armed with the findings of Section~\ref{sec:example}, we are now in a position to analyze the eigenvalue problem~(\ref{eq:bvp}) determining the magnetic buoyancy instability. Clearly, just as for equation~(\ref{eq:exODE}), the $k \to \infty$ limit is singular, with the coefficient of the highest (second) derivative (i.e.\  $\sigma^2 \rhobar^2 F(z) / \left( \sigma^2 + k^2 F(z) \right)$) tending to zero. This suggests that the eigenmodes become localized when $k$ becomes large and thus we may use singular perturbation techniques (boundary layer techniques) to solve the eigenvalue problem. For a certain eigenmode and eigenvalue $\sigma$ the main flow equation in the region where the derivatives are of order unity (i.e.\ outside any boundary/internal layer) is given, as in \cite{Gilman_1970}, by expression~(\ref{eq:Gilman}), i.e.\
\begin{equation}
\sigma^2 w = \frac{\Lambda \Bbar^2}{\rhobar F(z)}
\left( \frac{1}{H_\rho} - \frac{1}{H_B} \right) w .
\label{eq:Lklimitgrate}
\end{equation}
Since $\sigma$ is a constant, whereas $\Lambda \Bbar^2 \left( H_\rho^{-1} - H_B^{-1} \right) / \rhobar F(z)$ is a function of $z$, the only way to satisfy equation (\ref{eq:Lklimitgrate}) is for $w$ to be zero in the main flow, i.e.\ in the region where the $z$-derivatives of $w$ are negligible. Thus for $k\gg1$, we seek a mode associated with a specific eigenvalue $\sigma$ that is localized in the vicinity of $z=z_0$, defined as the value of $z$ where $\sigma^2 = \Lambda \Bbar^2 \left( H_\rho^{-1} - H_B^{-1} \right) / \rhobar F(z)$. (Another possibility is for the second derivative of $w$ to be large not only locally, but in a significant part of the fluid domain, i.e.\ that the solutions are strongly oscillatory; see Section~\ref{subsec:no_max}.) Hence we define a scaled length variable by
\begin{equation}
\xi=\frac{z-z_0}{\delta},
\label{eq:var_xi}
\end{equation}
where $\delta(k)$ is a measure of the thickness of the boundary/internal layer. It follows from introducing the new boundary layer variable into equation (\ref{eq:bvp}) and expanding all functions of $\xi$ and the growth rate $\sigma = \sigma_0 + \delta \sigma_1 + \delta^2 \sigma_2 + \ldots$ in powers of $\delta$, that the only distinguished limits possible require $\delta\sim k^{-2/(2+n)}$, with a balance between the second derivative and the terms proportional to $w$ of the form
\begin{subequations}\label{eq:possible_balances}
\begin{equation}
\left[\sigma_0^2 - \frac{\Lambda \Bbar^2\left( H_\rho^{-1} - H_B^{-1} \right)}{\rhobar F(z)}\right]w = \frac{\sigma_0^2} {\delta^2 k^2} 
\frac{\drm ^2 w}{\drm \xi^2} ,
\label{eq:possible_balances_1}
\end{equation}
\begin{equation}
\left[2\sigma_0\sigma_1-\xi\frac{\drm }{\drm z}
\left( \frac{\Lambda \Bbar^2 \left( H_\rho^{-1} - H_B^{-1} \right)}
{\rhobar F(z)}\right)\right]w = \frac{\sigma_0^2}{\delta^3 k^2}
\frac{\drm ^2 w}{\drm \xi^2} ,
\label{eq:possible_balances_2}
\end{equation}
\begin{equation}
\left[\sigma_1^2 + 2 \sigma_0 \sigma_2 -
\frac{1}{2} \xi^2 \frac{\drm ^2}{\drm z^2}
\left(\frac{\Lambda \Bbar^2 \left( H_\rho^{-1} - H_B^{-1} \right)}
{\rhobar F(z) }\right)\right]w
\nonumber
\end{equation}
\begin{equation}
\qquad \qquad \qquad \qquad \qquad \qquad
=\frac{\sigma_0^2}{\delta^4 k^2}
\frac{\drm ^2 w}{\drm \xi^2} , \quad \textrm{etc.},
\label{eq:possible_balances_3}
\end{equation}
\end{subequations}
where all the functions of $z$ are evaluated at $z=z_0$.

It is important to realize that in equations (\ref{eq:possible_balances}) we have neglected higher order terms; in particular, terms of $\mathrm{O}(\xi\delta)$ in (\ref{eq:possible_balances_1}), of $\mathrm{O}(\xi^2\delta)$ in (\ref{eq:possible_balances_2}) and of $\mathrm{O}(\xi^3\delta)$ in (\ref{eq:possible_balances_3}), which are, of course, negligible only if $\xi$ is not too large. The fact, that these terms become large and non-negligible as $\xi$ becomes comparable with $\delta^{-1}$ means that the perturbation problem is singular and that perturbation series such as $\sigma = \sigma_0 + \delta \sigma_1 + \delta^2\sigma_2 + \ldots$ are strictly asymptotic; thus we limit ourselves to calculation of the few first terms of the series.

Starting with $\sigma=\sigma_0+\mathrm{o}(1)$ and $\delta = k^{-1}$ we obtain
\begin{equation}
\frac{\drm ^2w}{\drm \xi^2} -
\left[1-\frac{\Lambda \Bbar^2}{\sigma_0^2\rhobar  F}
\left(\frac{1}{H_\rho} - \frac{1}{H_B} \right) \right] w = 0,
\label{eq:IntLeq_1}
\end{equation}
where, again, $H_\rho$, $H_B$, $\Bbar$, $\rhobar$ and $F(z)$ are evaluated at $z=z_0$. It follows that the only continuous solution with continuous derivatives is $w = \textrm{const.} =W$, with $W$ independent of $\xi$ and
\begin{equation}
\sigma_0^2 = \left. \frac{\Lambda \Bbar^2}{\rhobar F}
\left( \frac{1}{H_\rho} - \frac{1}{H_B} \right) \right|_{z=z_0}.
\label{eq:sigma0}
\end{equation}
Internal layers must then be introduced in order to match this $w =  \textrm{const.}$ solution to $w=0$ in the main flow. Importantly, in the analysis that follows, we assume that the basic state is such that $\sigma_0^2 > 0$, i.e.\ the system is unstable.

At this stage it is important to distinguish between various cases requiring different treatments; we consider these in turn below.  In Section~\ref{subsec:max_in_interior} we examine the case where $\sigma$, defined by equation~(\ref{eq:Gilman}), has a local maximum at $z_{\maxrm}$, with $0 < z_{\maxrm} < 1$; i.e.\ the growth rate is maximized strictly within the layer. Section~\ref{subsec:max_at_bdry}  considers the case where the growth rate is maximized at the boundary ($z=0$ or $z=1$), analyzing separately the generic case, when to leading order $\sigma$ varies linearly with $z$, and the special case when a true maximum of $\sigma$ happens to occur at the boundary. If we consider the evolution of the instability in terms of an initial value problem then, ultimately, the mode of maximum growth rate will prevail. That said, other modes may be significant at earlier times, depending on the initial perturbation. Thus in Section~\ref{subsec:no_max} and the Appendix we explore the large $k$ asymptotic solutions about a general point in the layer where the growth rate is not maximized.

\subsection{The most unstable modes when $z_0=z_{\maxrm}$}
\label{subsec:max_in_interior}

Here we are interested in the case where the function $\sigma(z)$ defined by equation~(\ref{eq:sigma(z)}) has a quadratic maximum at $z=z_{\maxrm}$, with $0 < z_{\maxrm} < 1$. Thus $\sigma_1 = 0$ and the point $z = z_{\maxrm}$ is surrounded by a layer of thickness $k^{-1/2}$; thus $\delta^{\prime\prime} = k^{-1/2}$. On introducing $\varsigma = (z -z_0) / \delta^{\prime\prime}$ and $\sigma = \sigma_0 + \delta^{\prime\prime^2} \sigma_2 + \mathrm{o}(\delta^{\prime\prime^2})$ (with $\sigma_0$ given by (\ref{eq:sigma0})), equation~(\ref{eq:possible_balances_3}) takes the following form:
\begin{equation}
\frac{\drm^2 w}{\drm \varsigma^2} - \left[ 2 \frac{\sigma_2}{\sigma_0} - \frac{1}{2} \varsigma^2 \Upsilon \right] w = 0,
\label{eq:IntL_max_eq}
\end{equation}
where
\begin{equation}
\Upsilon=\frac{1}{\sigma_0^2} \left. \frac{\drm^2 \sigma^2}{\drm z^2} \right|_{z=z_0}<0,
\label{eq:UPSILON_Lmax}
\end{equation}
and where the requirement that $\Upsilon<0$ comes from the assumption that we are considering a maximum (rather than a minimum) of $\sigma_0$ at $z=z_{\maxrm}$. Equation~(\ref{eq:IntL_max_eq}) is readily transformed into the parabolic cylinder equation~(\ref{eq:exPC}) by introducing
\begin{equation}
x = \left( -2 \Upsilon \right)^{1/4} \varsigma, \qquad a=\frac{\sigma_2}{\sigma_0} \sqrt{-\frac{2}{\Upsilon}} \, .
\label{eq:chv_z_max}
\end{equation}
Hence the modes localized in the vicinity of $z_{\maxrm}$ take the same form as in the example problem of Section~\ref{sec:example}, with $\sigma_2$ determined by the second relation in (\ref{eq:chv_z_max}) and by the fact that $a=-n-1/2$ with $n\in\mathbb{N}\cup \{0\}$.

\subsection{The most unstable modes when $z_0 =0$ or $1$} \label{subsec:max_at_bdry}

In the case when the maximum of the function $\sigma(z)$ given by (\ref{eq:sigma(z)}) lies strictly outside the layer, or when the function has no maximum, the fastest growing mode has growth rate $\sigma$ given by its value on the upper or lower boundary, depending whether $\sigma(z)$ is an increasing or decreasing function. Furthermore, since now $\sigma (z-z_0)$ varies linearly in $z-z_0$ (not quadratically as in Section~\ref{subsec:max_in_interior}) then this implies a different scaling for the boundary layer.

Here we take $\delta^{\prime} = k^{-2/3}$, $\zeta=(z-z_0)/\delta^{\prime}$, with $z_0=1$ or $z_0=0$ and $\sigma = \sigma_0 + \delta^{\prime} \sigma_1 + \mathrm{o}(\delta^{\prime})$ (with $\sigma_0$ given by expression~(\ref{eq:sigma0})), which, by the use of equation~(\ref{eq:possible_balances_2}), leads to
\begin{equation}
\frac{\drm^2 w} {\drm \zeta^2} - \left( 2 \frac{\sigma_1}{\sigma_0} -\zeta \Sigma \right)  w = 0,
\label{eq:IntLeq_2}
\end{equation}
where
\begin{equation}
\Sigma = \frac{1}{\sigma_0^2} \left. \frac{\drm \sigma^2}{\drm z} \right|_{z=z_0}.
\label{eq:SIGMA}
\end{equation}
On introducing the new variable $s = \left( 2\sigma_1/\sigma_0 - \zeta \Sigma \right) \Sigma^{-2/3}$, equation~(\ref{eq:IntLeq_2}) becomes Airy's equation,
\begin{equation}
\frac{\drm^2 w}{\drm s^2} - s w = 0,
\label{eq:Besseleq_1}
\end{equation}
with solution \citep{AS_1972},
\begin{equation}
w=C_{1}\mathrm{Ai}(s) + C_2\mathrm{Bi}(s),
\label{eq:General_SolB_1}
\end{equation}
where $C_{1}$ and $C_{2}$ are complex constants.

To fix ideas, let us assume that the function $\sigma^2(z)$ is an increasing function of $z$ at $z=z_0$, i.e.\ $\Sigma>0$; thus we concentrate on $z_0 = 1$. The solution in the region $\zeta<0$ takes the form
\begin{align}
w =
C_1 \mathrm{Ai} \left[ \frac{1}{\Sigma^{2/3}}
\left( 2 \frac{\sigma_1}{\sigma_0} - \zeta \Sigma \right) \right] &+ \nonumber \\ 
C_2 \mathrm{Bi} \Bigg[ \frac{1}{\Sigma^{2/3}} &
\left( 2 \frac{\sigma_1}{\sigma_0} - \zeta \Sigma \right) \Bigg] .
\label{eq:General_SolB_2}
\end{align}
Matching with the main flow solution requires that  $w(\zeta \to -\infty)=0$. Since $ \mathrm{Ai} (s)$ decays exponentially and $ \mathrm{Bi} (s)$ grows exponentially as $s \to \infty$, it follows that $C_2$ must be zero and hence
\begin{equation}
w = C_1 \mathrm{Ai} (s) =
C_1 \mathrm{Ai} \left[ \frac{1}{\Sigma^{2/3}}
\left( 2 \frac{\sigma_1}{\sigma_0} - \zeta \Sigma \right) \right].
\label{eq:uz_sol_L2_Hankel}
\end{equation}
Furthermore, solution (\ref{eq:uz_sol_L2_Hankel}) must also satisfy the impermeability condition at the upper boundary, i.e.\ $w(z=1)=0$. Thus
\begin{equation}
\mathrm{Ai} \left(  \frac{2}{\Sigma^{2/3}}
\frac{\sigma_1}{\sigma_0} \right) = 0 ,
\label{eq:Constants_L2_2}
\end{equation}
which determines $\sigma_1$ as
\begin{equation}
\sigma_1 = \frac{\mathcal{Z} \sigma_0 \Sigma^{2/3}}{2} ,
\label{eq:sig1_final_value}
\end{equation}
where $\mathcal{Z}<0$ denotes a zero of $\mathrm{Ai}(z)$ (all of the zeros of $\mathrm{Ai}(z)$ lie on the negative real axis, the smallest in magnitude being $-2.33811$).

For completeness, let us also consider the special case when the maximum of the function $\sigma(z)$ in (\ref{eq:sigma(z)}) is located at the boundary, say the upper boundary, $z_0=1$. The thickness of the boundary layer in this case is $\Orm (k^{-1/2})$; we thus take $\delta^{\prime\prime} \sim k^{-1/2}$ and introduce $\varsigma = (1-z)/\delta^{\prime\prime} > 0$ and $\sigma = \sigma_0 + \delta^{\prime\prime^2} \sigma_2 + \mathrm{o} (\delta^{\prime\prime^2})$ (with $\sigma_0$ given by (\ref{eq:sigma0}) and $\sigma_1=0$). By the use of (\ref{eq:possible_balances_3}) we obtain
\begin{equation}
\frac{\drm^2 w}{\drm \varsigma^2} - \left( 2 \frac{\sigma_2}{\sigma_0} - \frac{1}{2} \varsigma^2 \Upsilon \right) w = 0,
\label{eq:IntLeq_3}
\end{equation}
where
\begin{equation}
\Upsilon=\frac{1}{\sigma_0^2} \left. \frac{\drm^2 \sigma^2}{\drm z^2} \right|_{z=1}<0,
\label{eq:UPSILON}
\end{equation}
so that $\sigma(z)$ has a maximum at $z=1$. The solution is most easily expressed in terms of the parabolic cylinder function \citep{AS_1972}
\begin{equation}
w = C D_{-a-1/2} \left[\left( -2 \Upsilon \right)^{1/4} \varsigma \right],
\label{eq:uzsol_L3}
\end{equation}
where
\begin{equation}
a = \frac{\sigma_2}{\sigma_0} \sqrt{-\frac{2}{\Upsilon}},
\label{eq:L3_constant_a}
\end{equation}
$C$ is a constant, and where the second linearly independent solution has been omitted since it blows up exponentially in the main flow. The matching condition with the main flow, $w(\varsigma \to +\infty)=0$,
is satisfied since the asymptotic expansion of (\ref{eq:uzsol_L3}) for $\varsigma\gg1$ takes the form
\begin{equation}
w \sim C \left( -\frac{1}{2\Upsilon} \right)^{\frac{a+1/2}{4}} \frac{1}{\varsigma^{(a+1/2)}} \times 
\exp \Bigg[ -  \left( -\frac{\Upsilon }{8} \right)^{1/2} \varsigma^2 \Bigg] .
\label{eq:asymptform_L3}
\end{equation}
The impermeability condition at the upper boundary gives
\begin{equation}
D_{-a-1/2} \left(0 \right ) = 0 ,
\label{eq:constants_L3}
\end{equation}
which yields a condition for the permitted values of the parameter $a$ and hence for the correction to the growth rate, $\sigma_2$. Expression~(\ref{eq:constants_L3}) can be written as
\begin{equation}
D_{-a-1/2}(0) = \frac{\sqrt{\pi}}{2^{(2a+1)/4}\Gamma\left(\frac{3}{4}+\frac{1}{2}a\right)} = 0 ,
\label{eq:Uprime_expansion}
\end{equation}
where the requirement that the Gamma function is infinite implies $a=-2n-3/2$, where $n \in \mathbb{N} \cup \{0\}$. Taking $a=-3/2$, i.e.\ the smallest absolute value, gives
\begin{equation}
\sigma_{2} = -\frac{3}{2} \sigma_0 \sqrt{-\frac{\Upsilon}{2}} .
\label{eq:sig2_final_expr}
\end{equation}

\subsection{The modes at $z_0\neq z_{\maxrm}$ within the layer} \label{subsec:no_max}

The preceding analysis has demonstrated that in the $k\gg 1$ regime the fastest growing modes are strongly localized. Although in any temporal (and linear) evolution the mode of maximum growth rate, localized at  $z_0= z_{\maxrm}$, will ultimately dominate, initial conditions may be such that other modes prevail for some time. Thus it is of some interest also to consider the general case of eigenmodes associated with growth rates not in the vicinity of any extremum of the function $\sigma(z)$ defined by expression~(\ref{eq:sigma(z)}). In this case, however, it is not possible to produce a boundary layer solution varying strongly only in the locality of an isolated $z_0$ determined by $\sigma = \sigma(z_0)$. Instead, the eigenmodes exhibit strong oscillations throughout the bulk of the layer, a possibility that was noted earlier. This behavior can be captured only through a WKB (physical optics) analysis, which is more general, but considerably more complicated than the analysis of Sections~\ref{subsec:max_in_interior} and \ref{subsec:max_at_bdry}. For completeness we have included the detailed analysis for this case in the Appendix.

\section{Time Evolution}\label{sec:TE}

In Section~\ref{sec:MBE} we addressed the problem of magnetic buoyancy instability as an eigenvalue problem for the growth rate $\sigma$. Here we consider a different approach, solving the governing equations~(\ref{eq:pert_v}) -- (\ref{eq:pert_p}) as an initial value problem, in order to illuminate further what the concept of a growth rate that varies with height means for a physical system. For any given basic state we can determine the growth rate function $\sigma(z)$ in the $k \to \infty$ limit via equation~(\ref{eq:Gilman}). This can then be used to predict the long-term behavior of the perturbation variables for large, finite wavenumbers. Suppose we denote the maximum value of $\sigma$ in the range $0\leq z\leq 1$ by $\sigma_{\maxrm}$, attained at $z=z_{\maxrm}$. We study the temporal evolution of the perturbation variables at high  wavenumber ($k=1000$), with the aim of determining two things --- first, whether the eigenfunctions (and in particular, $w$) become localized around $z_{\maxrm}$, and, second, whether they ultimately grow at a rate close to $\sigma_{\maxrm}$. We solve four different problems in order to demonstrate all aspects of the system. Adopting $iv$ as one of the variables, the problem can be solved, without loss of generality, purely in terms of real variables \citep[see][]{HC_1987}.

Equations~(\ref{eq:pert_v}) -- (\ref{eq:pert_p}) (in which we assume $\alpha=1$ for simplicity and replace the growth rate $\sigma$ with $\partial/\partial t$) are solved using a second-order Adams-Bashforth timestep, employing second-order finite differences to approximate spatial derivatives. Although such a timestepping scheme does introduce numerical diffusion, the solutions have proven to be of a satisfactory quality. We impose impermeability boundary conditions, $w=0$ at $z=0$ and $z=1$. Growth rates are calculated by analyzing the growth of the energy in $w$, measured by the integral of $|w|^2/2$ across the layer.

\begin{figure}
\plotone{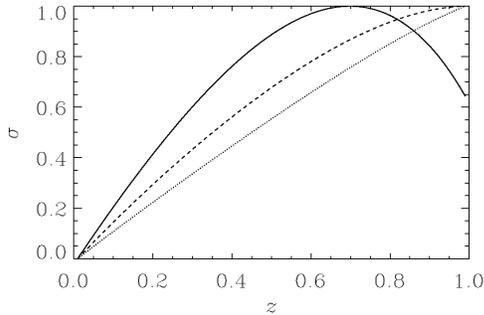}
\caption{Growth rate function $\sigma(z)$, normalized by its maximum value, versus $z$ for three sets of parameters. Case~1 (solid line), with  $\Lambda=0.2$, $\lambda=1.35$, $\mathcal{P}=1.9$; Case~2 (dotted line), with $\Lambda=0.3$, $\lambda=1.35$, $\mathcal{P}=1.0$; Case~3 (dashed line), with $\Lambda=0.2$, $\lambda=1.35$, $\mathcal{P}=1.48$.}
\label{fig:theorsigma}
\end{figure}

To cover the three main cases as described in Section~\ref{sec:MBE}, three different basic state configurations are considered --- specifically, the cases with $z_\maxrm$ located strictly within the layer, $z_\maxrm$ located at the boundary with $\sigma'(z_{\maxrm})$ non-zero, and $z_\maxrm$ located at the boundary with $\sigma'(z_{\maxrm}) = 0$. For the basic state, we adopt a linearly stratified magnetic field, of the form $\Bbar(z) = 1+\lambda (1-z)$. Case~1 has parameter values $\Lambda=0.2$, $\lambda=1.35$ and $\mathcal{P}=1.9$, and has maximal growth rate $\sigma_\maxrm=0.310$ at $z_\maxrm=0.698$; Case~2 has $\Lambda=0.3$, $\lambda=1.35$, $\mathcal{P}=1.0$, with $\sigma_\maxrm=0.417$ at $z_\maxrm=1.0$; Case~3 has $\Lambda=0.2$, $\lambda=1.35$, $\mathcal{P}=1.48$, with $\sigma_\maxrm=0.319$ at $z_\maxrm=1.0$. The three graphs of $\sigma$ against $z$ (scaled to be displayed on a single set of axes) are shown in Figure~\ref{fig:theorsigma}.

The first calculation is simply intended to show that, in the long term, the solutions do indeed peak at $z_\maxrm$ and that they grow with a rate close to $\sigma_\maxrm$. We adopt the basic state of Case~1, implying that the growth rate is peaked within the layer; we would thus expect the solutions eventually to be localized close to $z=0.698$. The initial conditions are an `educated guess' based on the results of \citet{HC_1987} for an $O(1)$ value of $k$; $iv=-0.02\cos(\pi z)$, $w=-0.07\sin(\pi z)$, $b_x=-\sin(\pi z)$ and $\rho=0.005\sin(\pi z)$.

\begin{figure}
\plotone{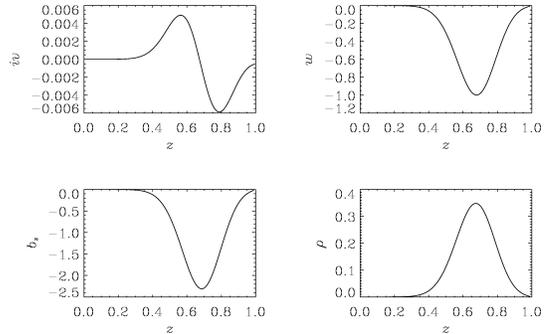}
\caption{Variables $iv$, $w$, $b_x$ and $\rho$ after $1000$ time units, plotted as functions of $z$; $k=1000$, $\Lambda=0.2$, $\lambda=1.35$, $\mathcal{P}=1.9$. The functions are normalized so that $\max(|w|)=1$.}
\label{fig:allefns}
\end{figure}

The spatial forms of the velocity, magnetic field and density (normalized so that $\max(|w|)=1$) after $1000$ time units are shown in Figure~\ref{fig:allefns}. The vertical velocity $w$ is peaked at $z=0.677$, close to the $k\to\infty$ value of $z_{\maxrm}=0.698$. The growth rate (calculated at the end of the simulation to ensure that the initial conditions have minimal effect) is $0.309$, also extremely close to the predicted value. In addition to $w$ being peaked close to $z=z_\maxrm$, we see that $b_x$ and $\rho$ are also peaked at $z \approx z_\maxrm$, with $v$ (which is out of phase with the other variables) going through zero.

\begin{figure}
\plotone{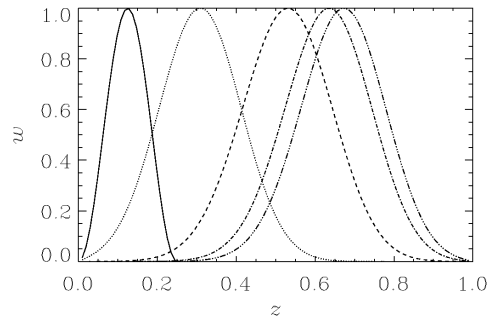}
\caption{Vertical velocity $w$ (normalized so that $\max(|w|)=1$) plotted as a function of $z$ at $t=0$ (leftmost plot), $t=750$, $t=1500$, $t=2250$ and $t=3000$ (rightmost plot); $k=1000$, $\Lambda=0.2$, $\lambda=1.35$, $\mathcal{P}=1.9$.}
\label{fig:winternal}
\end{figure}

A second calculation was then performed using the same parameter values but different initial conditions. For such a problem, we anticipate that the initial conditions should have no bearing on the final result, since the fastest growing mode should ultimately come to dominate. Initially, $w$ is sharply peaked at $z=0.12$, so that its peak is displaced far from the region in which we predict the solutions will eventually localize. Figure~\ref{fig:winternal} clearly shows the migration of the peak of the solution for $w$ towards $z=z_{\maxrm}$; after $3000$ time units the peak is located at $z=0.677$. The growth rate (calculated at the end of the simulation) is $0.310$, indicating that the initial conditions do not affect the ultimate growth rate, which is indeed determined by the basic state parameters.

\begin{figure}
\plotone{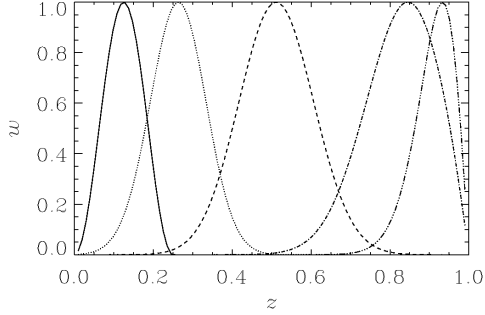}
\caption{Normalized vertical velocity $w$ at $t=0$ (leftmost plot), $t=250$, $t=500$, $t=750$ and $t=1000$ (rightmost plot); $k=1000$, $\Lambda=0.3$, $\lambda=1.35$, $\mathcal{P}=1.0$.}
\label{fig:wexternal}
\end{figure}

We now turn our attention to Case~2, in which the true maximum of the growth rate function $\sigma(z)$ is located outside the region $0<z<1$; within the layer, the growth rate is thus maximized at the boundary (in this case at $z=1$). In this situation there is a conflict between satisfying the boundary condition and the fact that the eigenfunction wishes to localize at this point. We again start with initial conditions in which $w$ is peaked at $z=0.12$. Figure~\ref{fig:wexternal} shows the peak of the vertical velocity $w$ moving towards the upper boundary and then becoming increasingly localized with time. As predicted by the analysis of Section~\ref{subsec:max_at_bdry}, the solution in this case adopts the form of an Airy function. Owing to the conflicting effects of the boundary condition and the system's desire to localize the eigenfunctions at the boundary, this is a more challenging numerical problem --- for this reason the calculation was continued for only $1000$ time units. The growth rate of the system is $0.408$, which is still close to $\sigma_{\maxrm}$ but is less accurate than for the calculation of Case~1; after $1000$ time units, the peak of $w$ is located at $z=0.931$.

\begin{figure}
\plotone{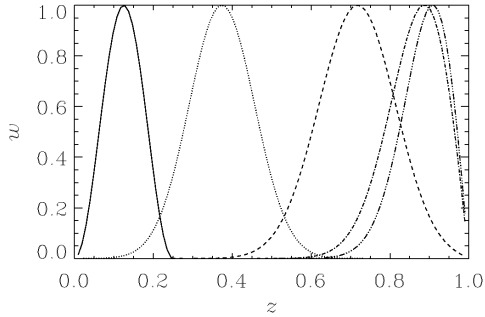}
\caption{Normalized vertical velocity $w$ at $t=0$ (leftmost plot), $t=500$, $t=1000$, $t=1500$ and $t=2000$ (rightmost plot); $k=1000$, $\Lambda=0.2$, $\lambda=1.35$, $\mathcal{P}=1.48$.}
\label{fig:wedge}
\end{figure}

Finally we consider Case~3, with $\sigma(z)$ truly maximized on the boundary (i.e.\ $\sigma(z)$ has vanishing derivative at $z=1$). In this case we expect the localized solution to be a parabolic cylinder function of the type displayed in Figure~\ref{fig:f_eigenmode}(b). Again, there is a competition between satisfaction of the homogeneous boundary condition and the system's desire to localize the eigenfunction at the boundary. Starting with the initial condition peaked at $z=0.12$, the system was evolved for $2000$ time units. The evolution of the solution towards the eigenfunction for $w$ can be seen in Figure~\ref{fig:wedge}; the growth rate towards the end of the simulation is $\sigma=0.316$, with the solution peaked at $z=0.912$.

\section{The 3D case}\label{sec:3D}

In this section we extend our earlier analysis so as to consider three-dimensional perturbations to the basic state described in Section~\ref{sec:Math_form}. We first show how the ideas of localized solutions developed for two-dimensional interchange modes carry over to the more general case. We then go on to elucidate the conditions under which two- or three-dimensional modes are preferred.

\subsection{Asymptotic analysis of 3D modes}\label{sec:3dasymp}

We now consider fully 3D perturbations of the form
\begin{subequations}\label{eq:pert_form_3D}
\begin{equation}
\bfu=\left(u\left(z\right),v\left(z\right),w\left(z\right)\right)\textrm{e}^{\sigma t}\textrm{e}^{i (k_{x} x+k_{y} y)},
\end{equation}
\begin{equation}
\bfb=\left(b_x\left(z\right),b_y\left(z\right),b_z\left(z\right)\right)\textrm{e}^{\sigma t}\textrm{e}^{i (k_{x} x+k_{y} y)},
\end{equation}
\begin{equation}
p=\tilde{p}\left(z\right)\textrm{e}^{\sigma t}\textrm{e}^{i (k_{x} x+k_{y} y)},\qquad\rho=\tilde{\rho}\left(z\right)\textrm{e}^{\sigma t}\textrm{e}^{i (k_{x} x+k_{y} y)}.\end{equation}
\end{subequations}
Introducing such perturbations into equations (\ref{eq:NS})--(\ref{eq:EQSTATE}) and linearizing about the basic state given by expressions~(\ref{eq:BS_field}), (\ref{eq:isotherm_BS_eq_2}) and (\ref{eq:BS_rho}), leads to the following set of equations:
\begin{equation}
\sigma \rhobar u=-i k_x\mathcal{P}\tilde{p}+\Lambda \Bbar' b_z,\label{eq:Pert3D_u}
\end{equation}
\begin{equation}
\sigma \rhobar v=-i k_y\left(\mathcal{P}\tilde{p}+\Lambda \Bbar b_x\right)+i k_x \Lambda \Bbar b_y,\label{eq:Pert3D_v}
\end{equation}
\begin{equation}
\sigma \rhobar w=-\frac{\drm}{\drm z}\left(\mathcal{P}\tilde{p}+\Lambda \Bbar b_x\right)+i k_x \Lambda \Bbar b_z-\tilde{\rho},\label{eq:Pert3D_w}
\end{equation}
\begin{equation}
\sigma b_x=-\left(i k_y v+\frac{\drm w}{\drm z}\right)\Bbar -\Bbar'  w,\label{eq:Pert3D_a}
\end{equation}
\begin{equation}
\sigma b_y=i k_x \Bbar v,\qquad \sigma b_z=i k_x \Bbar w,\label{eq:Pert3D_bc}
\end{equation}
\begin{equation}
\sigma\tilde{\rho}=-\rhobar \left(i k_x u+ i k_y v\right)-\frac{\drm}{\drm z}\left(\rhobar w\right),\label{eq:Pert3D_mass}
\end{equation}
\begin{equation}
0= i k_x b_x + i k_y b_y + \frac{\drm b_z}{\drm z},\label{eq:Pert3D_solB}
\end{equation}
\begin{equation}
\tilde{p}=\alpha\tilde{\rho}.\label{eq:Pert3D_state}
\end{equation}
Taking equations~(\ref{eq:Pert3D_u}) -- (\ref{eq:Pert3D_state}),  manipulating them to produce a single second-order ODE for $w$, and then applying the $k_y\gg1$ limit, yields (cf.\ equation~(\ref{eq:IntLeq_1}))
\begin{equation}
\frac{\drm^2w}{\drm z^2} - k_y^2(A_1A_2)^{-1} \left( \sigma^4 + A_1A_3\sigma^2 + A_1A_4 \right) w = 0,
\label{eq:3DODE}
\end{equation}
where
\begin{align*}
A_1 &= \frac{\Lambda\Bbar^2}{\alpha\calP\rhobar+\Lambda\Bbar^2},\\
A_2 &= \left(\frac{\sigma^2\rhobar}{\Lambda\Bbar^2}+k_x^2\right)\left(\sigma^2\left[\alpha\calP+\Lambda\frac{\Bbar^2}{\rhobar}\right]+\alpha\calP\Lambda\frac{\Bbar^2}{\rhobar}\right),\\
A_3 &= k_x^2\Lambda\frac{\Bbar^2}{\rhobar}\left(2\frac{\alpha\calP}{\Lambda}\frac{\rhobar}{\Bbar^2}+1\right)+\frac{1}{H_B}-\frac{1}{H_\rho}, \\
A_4 &= k_x^2\Lambda\frac{\Bbar^2}{\rhobar}\left(\alpha\calP k_x^2+\frac{1}{H_B}\right).
\end{align*}
Note that the term proportional to the first derivative in (\ref{eq:3DODE}) has been neglected, since, as in the 2D case, it is always much smaller than the second derivative term, which must be of comparable magnitude to the term proportional to $w$, at least in part of the domain. 

As for the interchange modes, letting $k_y \to \infty$ yields a purely algebraic equation for the growth rate, namely
\begin{equation}
\sigma^4 + A_1 A_3 \sigma^2 + A_1 A_4 = 0.
\label{eq:3d_disp}
\end{equation}
Alternatively, this can be derived by writing $\xi=(z-z_0)/\delta$, with $\delta\sim k_y^{-1}$, $\sigma=\sigma_0+\orm (1)$, and evaluating the functions $A_1$, $A_2$, $A_3$ and $A_4$ at $z=z_0$. Equation~(\ref{eq:3d_disp}) is of biquadratic form and hence gives two solutions, $\sigma_{+}$ and $\sigma_{-}$, say, for the square of the `depth-dependent growth-rate function',  where the subscript indicates which sign is taken in the quadratic formula. Written explicitly, the two roots are
\begin{equation}
\sigma^2_{0,\pm} = -\frac{A_1A_3}{2} \pm \frac{1}{2} \left[ A_1 ( A_1A_3^2-4A_4 ) \right]^{1/2}. \label{eq:sigma3D}
\end{equation}
This defines two functions for the square of the growth rate. We will now assume that the maximal value of $\sigma$ is attained within the layer, at, say, $0 < z_{\maxrm} < 1$, which implies
\begin{equation}
\sigma^2 \frac{\drm}{\drm z} \left. \left(A_1 A_3 \right) \right|_{z=z_\maxrm} + 
\frac{\drm}{\drm z} \left. \left( A_1 A_4 \right) \right|_{z=z_\maxrm}=0;
\label{eq:maximum_grate_3D_condition}
\end{equation}
for now, it does not matter which root ($\sigma_{+}$ or $\sigma_{-}$) attains this value, which we will call $\sigma_{\maxrm}$. Following the analysis of Section~\ref{subsec:max_in_interior}, we seek solutions localized around $z_0=z_{\maxrm}$, where $\sigma_0=\sigma_{\maxrm}.$ As in the 2D analysis, we expect to find localized solutions on applying the scaling $\delta^{\prime\prime}\sim k_y^{-1/2}$. On writing $\varsigma=(z-z_0)/\delta^{\prime\prime}$, with $\sigma=\sigma_0+\delta^{\prime\prime2}\sigma_2+ \orm (\delta^{\prime\prime2})$, equation~(\ref{eq:3DODE}) yields
\begin{align}
&\frac{\drm^2w}{\drm\varsigma^2} - \frac{1}{\left.(A_1A_2)\right|_{z=z_0}} 
\bigg[ 2\sigma_0 \sigma_2\left( 2\sigma_0^2 + \left. \left( A_1 A_3 \right) \right|_{z=z_0} \right) + \bigg.\\
&+ \left. \frac{\varsigma^2}{2} \left( \sigma_0^2 \frac{\drm^2}{\drm z^2} \left. (A_1 A_3) 
\right|_{z=z_0} +\frac{\drm^2}{\drm z^2}\left.(A_1A_4)\right|_{z=z_0}\right)\right]w=0, \nonumber
\end{align}
which can be recast into the parabolic cylinder equation~(\ref{eq:exPC}) under the coordinate transformation
\begin{align*}
x &=\left[\frac{2(\sigma_0^2\frac{\drm^2}{\drm z^2}\left.(A_1A_3)\right|_{z=z_0}+\frac{\drm^2}{\drm z^2}\left.(A_1A_4)\right|_{z=z_0})}{\left.(A_1A_2)\right|_{z=z_0}}\right]^{1/4}\varsigma , \\
a &= \frac{\sigma_0\sigma_2(2\sigma_0^2+\left.(A_1A_3) \right|_{z=z_0})}
{\left. \left( A_1 A_2 \right)^{1/2} \right|_{z=z_0} }\times \\ 
&\ \ \ \ \  \frac{\sqrt{2}}{\left(\sigma_0^2\frac{\drm^2}{\drm z^2}\left.(A_1A_3)\right|_{z=z_0}+\frac{\drm^2}{\drm z^2}\left.(A_1A_4)\right|_{z=z_0}\right)^{1/2}} .
\end{align*}
It exhibits localized solutions in the case where
\begin{equation}
\sigma_0^2\frac{\drm^2}{\drm z^2}\left.(A_1A_3)\right|_{z=z_0}+\frac{\drm^2}{\drm z^2}\left.(A_1A_4)\right|_{z=z_0}
\end{equation}
is positive.

Thus the general ideas and analysis of the localized solutions explored in Section~\ref{subsec:max_in_interior} for 2D interchange perturbations carry through, albeit in a more involved fashion, to the general 3D case. Similarly, we anticipate that the 2D results of Sections~\ref{subsec:max_at_bdry} and \ref{subsec:no_max} could be extended to cover the 3D instability.

\subsection{Preferred mode of instability}\label{sec:preferred_mode}

Having derived the depth-dependent dispersion relations for 2D (interchange) and 3D perturbations, it is of interest to tie these together in order to determine the preferred mode of instability, and to validate them against previous magnetic buoyancy analyses. Writing the dispersion relation~(\ref{eq:3d_disp}) in full gives
\begin{align}
0 = &\left(\frac{\alpha\mathcal{P}}{\Lambda}\frac{\rhobar}{\Bbar^2}+1\right)\sigma^4+\left[k_x^2\Lambda \frac{\Bbar^2}{\rhobar}\left(2\frac{\alpha\mathcal{P}}{\Lambda}\frac{\rhobar}{\Bbar^2}+1\right)\right.\nonumber \\
& \bigg.+\frac{1}{H_B}-\frac{1}{H_\rho}\bigg]\sigma^2+k_x^2\Lambda \frac{\Bbar^2}{\rhobar}\left(\alpha \mathcal{P} k_x^2 +\frac{1}{H_B}\right).
\label{eq:3D_dispersion}
\end{align}
Since magnetic buoyancy instabilities are of interest only for bottom-heavy equilibria, we shall only consider the case of $H_\rho<0$. From the dispersion relation~(\ref{eq:3D_dispersion}) it can be shown that, as expected on physical grounds, instability sets in only as a direct mode, with $\sigma$ passing through zero. Instability to 3D modes occurs provided that \begin{equation}
 k_x^2<-\frac{1}{\alpha\mathcal{P }H_B}.
\label{eq:3D_instability}
\end{equation}
Thus long wavelength (in the $x$-direction) modes are favored, and these can be destabilized solely by a decrease in height of the magnetic field ($H_B<0$), rather than the decrease in height of $B/\rho$ required to destabilize interchange modes (equation~(\ref{eq:inst_criterion})). This simply confirms the original results of \citet{Newcomb_1961}, derived from the energy principle of ideal MHD. The physical explanation underlying why 3D modes can be more readily destabilized than interchange modes, despite having to do work against magnetic tension, is discussed at length in \citet{HC_1987}.

It is also of interest to determine the mode of maximum growth rate once the instability criterion (\ref{eq:3D_instability}) is satisfied. We find the wavenumbers $k_x$ at which the function $\sigma^2(k_x)$ achieves extremal values by solving $\drm \sigma^2/\drm k_x=0$. Calculation of the second derivative $\drm^2\sigma^2/\drm k_x^2$ then allows us to determine the positive extrema. By fairly simple algebraic manipulations, one can demonstrate that the extremal values of $\sigma^2(k_x)$ are achieved for $k_{0x}=0$ and $k_{1x}$, where the latter is defined by the real roots of the biquadratic equation
\begin{align}
0 = k_{1x}^4 &- \frac{2}{\Lambda}\frac{\rhobar}{\Bbar^2}\left[\frac{1}{H_B}+\frac{1}{H_\rho}\left(2\frac{\alpha\mathcal{P}}{\Lambda}\frac{\rhobar }{\Bbar^2}+1\right)\right]k_{1x}^2 \nonumber \\
& + \frac{1}{\Lambda^2H_B}\frac{\rhobar ^2}{\Bbar^4}\left[\frac{1}{H_B}-\frac{1}{H_\rho}\left(2+\frac{\Lambda}{\alpha\mathcal{P}}\frac{\Bbar^2}{\rhobar }\right)\right],\label{eq:k_at_extremum}
\end{align}
where all the functions are evaluated at a fixed $z=z_0$; thus $k_{1x}$ depends on height. For sufficiently strong field gradients, i.e.\ for
\begin{equation}
-\invHB>-\left(2+\chi\right)\invHrho,
\label{strong_Xi}
\end{equation}
where
\begin{equation}
\chi=\frac{\Lambda}{\alpha\mathcal{P}}\frac{\Bbar^2}{\rhobar },
\label{chi}
\end{equation}
the growth rate is maximized at $k_{0x}=0$ and
\begin{equation}
\sigma_{\maxrm}^2 = \frac{\Lambda \Bbar^2}{\rhobar \left(\alpha\mathcal{P} + \Lambda \Bbar^2/\rhobar \right)} \left(\frac{1}{H_\rho} - \frac{1}{H_B} \right),
\label{sigma_at_strong_Xi}
\end{equation}
in agreement with (\ref{eq:sigma0}). In other words, for sufficiently strong field gradients, interchange modes possess the largest growth rate. For $-\invHrho < -\invHB < -(2+\chi)\invHrho$, the growth rate $\sigma$ has a local minimum at $k_{0x} = 0$; in this range, $\sigma^2(k_x)$ is maximized at $k_x = k_{1x}$. For $0< -\invHB < -\invHrho$, $\sigma^2$ given by expression~(\ref{sigma_at_strong_Xi}) is no longer positive, but the maximum of $\sigma^2(k_x)$ at $k_{1x}$ remains positive. These findings are in keeping with those of \citet{Hughes_1985}, who examined magnetic buoyancy instabilities under the magneto-Boussinesq approximation.

Finally, since $\chi\sim (\alpha\beta)^{-1}$, we note that this parameter would become negligibly small in the parameter regime of stellar interiors ($\beta\gg1$, $\alpha\sim 1$). The above results then simplify to saying that for $-\invHB\gtrsim-2\invHrho$ the dominant perturbations are two-dimensional interchange modes (the growth rate has a maximum at $k_x=0$), whereas for $-\invHB \lesssim -2 \invHrho$, finite wavelengths in the $x$ direction are preferred (growth rate maximized at $k_{1x}$).

\section{Discussion}\label{sec:Disc}

Through the use of a Rayleigh-Schr\"odinger perturbation analysis to exploit the large wavenumber in the horizontal transverse direction, we have revisited the problem introduced in the pioneering works of \citet{Gilman_1970} and \cite{Acheson_1979} of the magnetic buoyancy instability in a plane layer of fluid permeated by an external horizontal magnetic field decreasing with height. Such an approach is simpler and more physically appealing than a WKB analysis, although is less general in its scope. We have studied in detail the instability of interchange modes (no bending of the field lines) and, consequently, have been able to provide a thorough mathematical explanation of the results of \citet{Gilman_1970}, who obtained instability growth rates as functions of the vertical coordinate $z$. Via an asymptotic analysis of the governing equations, we have shown how the fastest growing eigenmodes may become strongly localized in the vertical direction. The growth rate function $\sigma(z)$ is maximized either strictly within the layer (at $0 < z_\maxrm < 1$, say) or else at the boundary of the domain. In the former case, the mode of maximum growth rate is localized at $z_\maxrm$ and adopts the form of a parabolic cylinder function. In the latter case, the eigenfunction is localized close to the boundary and takes the form of an Airy function. The modes associated with growth rates that are less than maximal exhibit strong oscillations; these can be captured only via a WKB approach, as presented in the Appendix.

We have confirmed the results of the asymptotic analysis by numerical simulations of the time evolution of the perturbations, posed as an initial value problem. These clearly show that, starting from an initial condition composed of a set of eigenmodes, the mode that eventually dominates the evolution, in the large $k$ limit, is that with growth rate corresponding to the maximum of the growth rate function $\sigma(z)$. Furthermore, this mode is localized in the vicinity of $z_{\maxrm}$, the height in the layer  at which $\sigma(z)$ is maximized.

The asymptotic analysis carries over, albeit with more algebraic complications, to the case of three-dimensional perturbations that are infinitesimally thin in the horizontal direction perpendicular to the imposed field, but have a finite (though long) wavelength along the field. From an analysis of the dispersion relation~(\ref{eq:3D_dispersion}), we have shown how 3D perturbations are the only unstable modes for weakly unstable magnetic field gradients, but that for sufficiently strong field gradients, interchange modes have the largest growth rate.

It is of interest to consider what the analysis tells us about the specific physics of the magnetic buoyancy instability and, more generally, to enquire into the wider class of instabilities that can be so analyzed. Magnetic buoyancy instabilities are driven by a destabilizing magnetic field gradient and inhibited by a stabilizing entropy gradient. The latter is eroded by a combination of high thermal diffusivity and small wavelengths transverse to the imposed field; in the limiting case we have considered here, the thermal diffusion tends to infinity and the wavelength to zero. The crucial feature revealed by the analysis of Section~\ref{sec:MBE} is that the assumption of locality in the horizontal direction implies locality in the vertical. For a case when the growth rate function $\sigma(z)$ is positive in only a small section of the layer, it is perhaps not surprising that the eigenfunction of the fastest growing mode is peaked in that region; the important feature revealed here is that the fastest growing eigenfunction is always peaked where $\sigma(z)$ is maximized, even when $\sigma(z)$ is positive throughout the layer.

In a wider context, is it possible to identify, simply from general considerations, other instabilities that can be analyzed within the same framework? Certainly two necessary conditions can be identified: one is a physical reason for an instability to have a small wavelength in one specific direction; the other is an inhomogeneity of the basic state in an orthogonal direction. For magnetic buoyancy instability as discussed here, the small wavelength in the $y$-direction arises from the erosion of the stabilizing gradient through thermal diffusion, whereas the inhomogeneity in the vertical direction is a consequence of the $z$-dependence of the basic state magnetic field and thermodynamic variables. In a Boussinesq atmosphere, in which stratification is uniform with depth, we would therefore not expect any localization in the vertical --- this is consistent with the magneto-Boussinesq analysis of \cite{Hughes_1985}. For the very different problem of the instability of a rotating, stratified flow with arbitrary horizontal cross-stream shear, considered by \cite{Griffiths_2008}, the assumed small scale is in the vertical, resulting from the stable stratification, with the inhomogeneity resulting from the shear flow. 

In the present analysis we have neglected viscosity and magnetic diffusivity and assumed infinitely fast thermal diffusion. The effects of finite diffusion will of course influence the dynamics at sufficiently large $k$, and will act to establish a finite magnitude of the wave vector for unstable modes. Thermal diffusion will, however, still greatly dominate and thus, although consideration of the full dynamics will lead to modification of the quantitative results, we expect that the short-wavelength nature in the horizontal direction perpendicular to the field and the strong localization in the vertical direction of the most unstable perturbations to persist.

\acknowledgments
We are grateful to Douglas Gough, Stephen Griffiths, Chris Jones, Michael Proctor and Steve Tobias for valuable discussions. This work was funded by the STFC rolling grant held at the University of Leeds.

\appendix

\section{Modes with growth rates less than maximal}

For completeness, here we consider the structure of the solutions in the case where the growth rate $\sigma$ is not maximal in $0 \le z \le 1$. As explained in Section~\ref{subsec:no_max}, the boundary layer analysis of Sections~\ref{subsec:max_in_interior} and \ref{subsec:max_at_bdry} is not suitable in this case, and it is thus necessary to adopt a more general, but more complicated, WKB approach \citep[cf.][]{BenderOrszag}. We thus seek solutions of equation~(\ref{eq:bvp}) of the form
\begin{equation}
w=C\exp{\left[\frac{1}{\delta}\sum_{n=0}^{\infty}\delta^n S_n(z)\right]},
\label{eq:WKB_expansion}
\end{equation}
where $\delta \ll 1$ is a function of $k \gg 1$ (to be determined) and $C$ is a constant; $w$ satisfies the boundary conditions $w=0$ at $z=0$ and $z=1$. We will only consider the first two terms of the expansion (\ref{eq:WKB_expansion}), i.e.\ the `physical optics' approximation; this constitutes an asymptotic approximation to the solution of (\ref{eq:bvp}) since the subsequent terms are small provided the functions $S_n(z)$ are bounded. Introducing the expansion (\ref{eq:WKB_expansion}) into equation~(\ref{eq:bvp}) yields $\delta=k^{-1}$. We can then derive the following equations at zeroth and first orders:
\begin{subequations}\label{eq:WKB_hierarchy}
\begin{equation}
\frac{\drm S_0}{\drm z} = \pm\sqrt{1-\left(\frac{\sigma(z)}{\sigma}\right)^2} = \pm \sqrt{\calA(z)},\ \ \textrm{say,}
\end{equation}
\begin{equation}
0 = 2\left( \frac{\drm S_0}{\drm z} \right) \left( \frac{\drm S_1}{\drm z} \right) + \frac{\drm^2 S_0}{\drm z^2} + \frac{\drm S_0}{\drm z} \left( \frac{1}{F(z)} + \frac{2}{H_{\rho}} - \sigma(z)^2 \right).
\end{equation}
\end{subequations}
Let us first consider the case when $\sigma(z)$, defined by expression~(\ref{eq:sigma(z)}), is a monotonic function; without loss of generality we shall assume that it is increasing with $z$. For a given $z = z_0$ in $0 < z < 1$, the function $1-\sigma(z)^2/\sigma^2$ (henceforth denoted by $\calA(z)$) then has one zero within the layer; we therefore identify this as a one-turning-point WKB problem. We now divide the interval $[0,1]$ into three regions: region I defined by $z \geq 0$ and $z_0-z \gg k^{-2/3}$, region II by $|z-z_0| \ll 1$ and region III by $z \leq 1$ and $z-z_0 \gg k^{-2/3}$. From equations~(\ref{eq:WKB_expansion}) and (\ref{eq:WKB_hierarchy}), the general WKB solution in regions I and III takes the form
\begin{subequations}\label{eq:WKB_solutions_in_2_regions}
\begin{equation}
w_{I} = \left( \calA (z) \right)^{-1/4} \erm^{-\chi(z)} \left\{ C_1 \exp{\left( k \int_z^{z_0} \sqrt{\calA (z)} \drm z \right) } + C_2 \exp{ \left( -k \int_z^{z_0} \sqrt{\calA (z)} \drm z \right) } \right\},
\end{equation}
\begin{equation}
w_{III} = \left( -\calA (z) \right)^{-1/4} \erm^{-\chi(z)} \left\{ C_5 \sin {\left( k \int_{z_0}^z \sqrt{-\calA (z) } \drm z \right)} + C_6 \cos{\left( k \int_{z_0}^z \sqrt{-\calA(z)} \drm z \right) }\right\},
\end{equation}
\end{subequations}
where $C_1$, $C_2$, $C_5$ and $C_6$ are constants, the function $\chi(z)$ is defined by
\begin{equation}
\chi(z) = \frac{1}{2} \int \left( \frac{1}{F(z)} + \frac{2}{H_{\rho}} - \sigma(z)^2 \right) \drm z,
\end{equation}
and the span of regions I and III is determined by the regions of validity of the approximations~(\ref{eq:WKB_solutions_in_2_regions}a) and (\ref{eq:WKB_solutions_in_2_regions}b) respectively. On applying the boundary conditions, $w_I(0)=0$ and $w_{III}(1)=0$, we can rewrite (\ref{eq:WKB_solutions_in_2_regions}) in the form
\begin{subequations}\label{eq:WKB_gen_solutions_2regions}
\begin{equation}
w_{I}  = C_2 \left( \calA (z) \right)^{-1/4} \erm^{-\chi(z)} \exp{\left(-k \int_z^{z_0} \sqrt{\calA (z)}\drm z \right)}  \left\{1 - \exp{\left[ -2k \left( \int_0^{z_0} \sqrt{\calA (z)} \drm z - \int_z^{z_0} \sqrt{\calA (z)} \drm z \right) \right]}\right\},
\end{equation}
\begin{equation}
w_{III} = C_5 \left( - \calA (z) \right)^{-1/4} \erm^{-\chi(z)} \, \frac{ \sin{ \left[ k\left( \int_{z_0}^z \sqrt{-\calA (z)} \drm z - \int_{z_0}^1 \sqrt{-\calA(z)} \drm z \right)\right]}}{\cos{\left[k\int_{z_0}^1\sqrt{-\calA(z)} \drm z\right]}}.
\end{equation}
\end{subequations}
The factor inside the braces in (\ref{eq:WKB_gen_solutions_2regions}a) is approximately equal to unity provided $z_0\gg k^{-2/3}$; the exponential term will however be retained for clarity.

\begin{figure}
\plotone{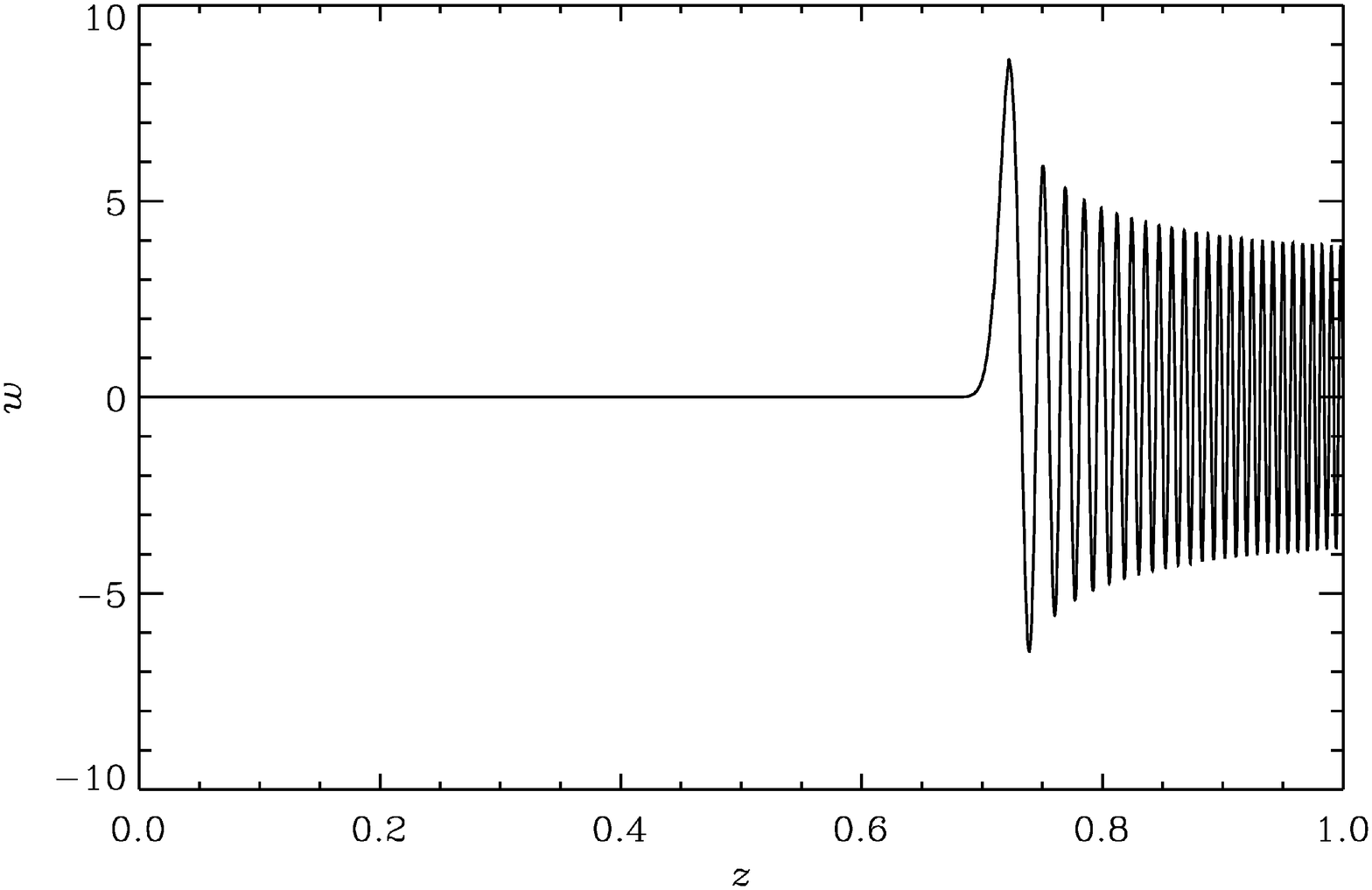}
\caption{Vertical velocity $w$ for the one-turning-point problem (Case~2 of Section~\ref{sec:TE}) with $C=1$, $k=2000$ and $n=100$.}
\label{fig:1tpproblem}
\end{figure}

Region II, defined by $|z-z_0|\ll1$, contains the turning point $z=z_0$ and thus the WKB approximation is not valid here. However, we can expand $\calA(z)$ as
\begin{equation}
\calA(z) = -\Sigma\left( z - z_0 \right) + \ldots, \ \textrm{ where }\ \Sigma = \frac{1}{\sigma^2} \left. \frac{\drm \left( \sigma(z)^2\right)}{\drm z}\right|_{z=z_0}>0,
\label{eq:expansion_of_the_grate}
\end{equation}
which leads to the differential equation
\begin{equation}
-\Sigma\left(z-z_0 \right) w = \frac{1}{k^2} \frac{\drm^2 w}{\drm z^2},
\label{eq:app_eq_regII}
\end{equation}
with solution expressed in terms of Airy functions,
\begin{equation}
w_{II} = C_3 \mathrm{Ai} \left[ k^{2/3} \Sigma^{1/3}(z_0-z)\right] + C_4 \mathrm{Bi} \left[k^{2/3} \Sigma^{1/3} (z_0-z) \right].
\label{eq:solution_regII}
\end{equation}
Matching between regions I and II requires the use of the asymptotic forms of Airy functions with large positive arguments \citep{AS_1972},
\begin{equation}
\mathrm{Ai}(\xi) \sim \frac{1}{2\sqrt{\pi}}\xi^{-1/4} \exp{\left( -2\xi^{3/2}/3 \right)} \quad \textrm{and} \quad
\mathrm{Bi}(\xi) \sim \frac{1}{\sqrt{\pi}} \xi^{-1/4} \exp \left( 2 \xi^{3/2}/3 \right),
\label{eq:Airy_asymptotic_large_xi}
\end{equation}
together with the asymptotic form of $w_I(z)$ for $0<z_0-z\ll1$ when (\ref{eq:expansion_of_the_grate}) holds,
\begin{equation}
w_{I} = C_2\frac{\erm^{-\chi(z_0)}}{\Sigma^{1/4}\left(z_0-z\right)^{1/4}}\left\{\erm^{-2k\Sigma^{1/2}(z_0-z)^{3/2}/3}-\erm^{2k\Sigma^{1/2}(z_0-z)^{3/2}/3}\exp{\left[-2k\int_0^{z_0}\sqrt{\calA(z)}\drm z\right]}\right\},
\label{eq:wI_asymptotic}
\end{equation}
leading to
\begin{equation}
C_3 = 2 \sqrt{\pi} \Sigma^{-1/6}k^{1/6} \erm^{-\chi(z_0)} C_2 \quad \textrm{and} \quad  C_4 = -\sqrt{\pi} \Sigma^{-1/6}k^{1/6} \erm^{-\chi(z_0)} \exp{ \left[ -2k \int_0^{z_0} \sqrt{\calA(z)} \drm z \right]} C_2 \ll C_2.
\label{eq:constants_234}
\end{equation}
On the other hand, matching the solutions in regions II and III requires use of the asymptotic forms of Airy functions with large negative arguments,
\begin{equation}
\mathrm{Ai}(\xi) \sim \frac{1}{\sqrt{\pi}} \left(-\xi\right)^{-1/4} \sin \left[\frac{2}{3} \left( - \xi \right)^{3/2} + \frac{\pi}{4} \right] \quad \textrm{and} \quad \mathrm{Bi}(\xi)\sim \frac{1}{\sqrt{\pi}} \left(-\xi\right)^{-1/4} \cos \left[\frac{2}{3} \left(-\xi\right)^{3/2} + \frac{\pi}{4}\right],
\label{eq:Airy_asympt_algebr}
\end{equation}
together with the asymptotic form of $w_{III}(z)$ for $0<z-z_0\ll1$ when (\ref{eq:expansion_of_the_grate}) holds,
\begin{equation}
w_{III} = C_5\frac{\erm^{-\chi(z_0)}}{\Sigma^{1/4}\left(z-z_0\right)^{1/4}}\frac{\sin{\left[\frac{2}{3}k\Sigma^{1/2}(z-z_0)^{3/2}-k\int_{z_0}^1\sqrt{-\calA(z)}\drm z\right]}}{\cos{\left[k\int_{z_0}^1 \sqrt{-\calA(z)}\drm z\right]}},
\label{eq:wIII_asymptotic}
\end{equation}
which, together with (\ref{eq:constants_234}), lead to
\begin{equation}
C_5 = \frac{\sqrt{2}}{2} \left\{ 2 + \exp \left[ -2k \int_0^{z_0} \sqrt{\calA (z)} \drm z \right]  \right\} C_2 \approx \sqrt{2}C_2,
\label{eq:constants_52}
\end{equation}
and
\begin{equation}
\sin\left[2k\int_{z_0}^1 \sqrt{-\calA(z)}\drm z\right]\approx -1.
\label{eq:eigenvalues_condition_sigma_increasing}
\end{equation}
Equation (\ref{eq:eigenvalues_condition_sigma_increasing}) is an eigenvalue condition. This completes the leading order WKB solution for the one-turning-point problem in the limit $k\gg1$ under the assumption that $\sigma(z)$ is an increasing function, i.e.\ $\Sigma > 0$ in the vicinity of $z=z_0$. As an illustration, Figure~\ref{fig:1tpproblem} shows the form of this solution for Case 2 of Section~\ref{sec:TE}, with $C=1$, $k=2000$, $n=100$ and (from equation~\ref{eq:eigenvalues_condition_sigma_increasing}) $z_0=0.716$.

\begin{figure}
\plotone{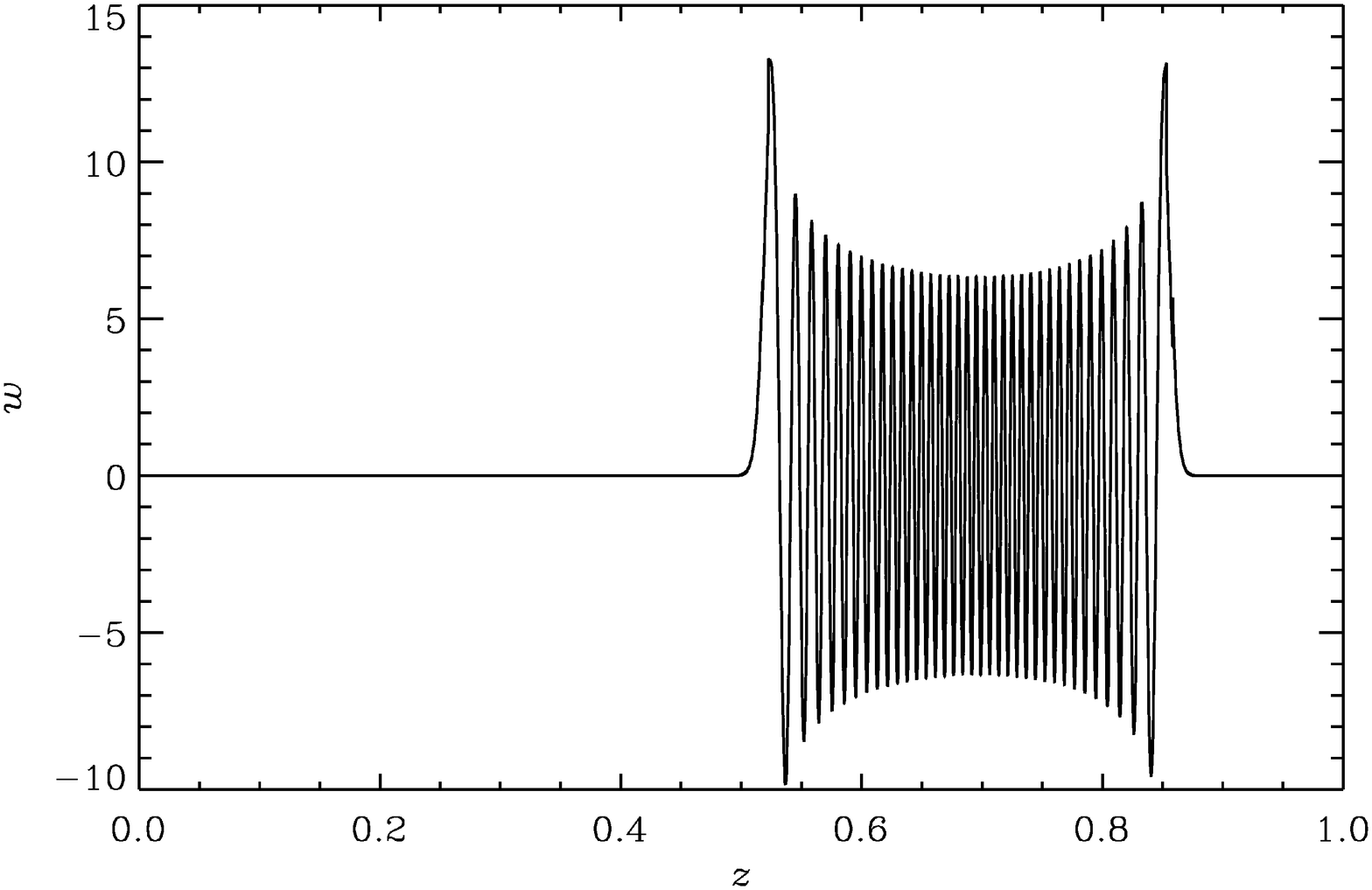}
\caption{Vertical velocity $w$ for the two-turning-point problem (Case~1 of Section~\ref{sec:TE})  with $C=1$, $k=8000$ and $n=70$.}
\label{fig:2tpproblem}
\end{figure}

We now turn our attention to the case where $\calA(z)$ has two zeros within the layer, at $z_1$ and $z_2>z_1$ defined by $\sigma=\sigma(z_1)=\sigma(z_2)$; this corresponds to a two-turning-point WKB problem. Additionally, let us suppose that $\calA(z)$ is positive for $z<z_1$ and $z>z_2$ and negative for $z_1<z<z_2$ (which corresponds to a concave function $\sigma(z)$ with a quadratic maximum between the points $z_1$ and $z_2$). The solution is obtained by asymptotic matching of two one-turning-point solutions \citep[cf.][]{BenderOrszag}. Since the analysis is very similar to that of the one-turning-point problem we simply give the approximate solution,
\begin{subequations}\label{eq:WKB_solutions_2TP_general}
\begin{equation}
w_{I} = C \left( \calA(z) \right)^{-1/4} \erm^{-\chi(z)} \exp \left( -k\int_z^{z_1} \sqrt{\calA(z)} \drm z \right) ,\qquad z \geq 0, \quad z_1 - z \gg k^{-2/3},
\end{equation}
\begin{equation}
w_{II} = 2\sqrt{\pi} C \Sigma_1^{-1/6} k^{1/6} \erm^{-\chi(z_1)} \mathrm{Ai} \left[k^{2/3} \Sigma_1^{1/3} (z_1-z) \right],\qquad\qquad \left|z-z_1\right|\ll1,
\end{equation}
\begin{equation}
w_{III} = 2 C \left(-\calA(z)\right)^{-1/4} \erm^{-\chi(z)} \sin {\left( k \int_{z_1}^z \sqrt{-\calA(z)} \drm z+\frac{\pi}{4}\right)},\qquad z-z_1\gg k^{-2/3},\quad z_2-z\gg k^{-2/3},
\end{equation}
\begin{equation}
w_{IV} = (-1)^n 2 \sqrt{\pi} C \left(-\Sigma_2\right)^{-1/6} k^{1/6} \erm^{-\chi(z_2)} \mathrm{Ai} \left[ k^{2/3} \left(-\Sigma_2 \right)^{1/3} (z-z_2) \right],\qquad\qquad \left|z-z_2\right|\ll1,
\end{equation}
\begin{equation}
w_{V} = (-1)^n C \left(\calA(z)\right)^{-1/4} \erm^{-\chi(z)} \exp{\left( -k \int_{z_2}^{z} \sqrt{\calA(z)} \drm z \right)},\qquad z \leq 1, \quad z-z_2 \gg k^{-2/3},
\end{equation}
\end{subequations}
where $n\geq0$ is a nonnegative integer,
\begin{equation}
\Sigma_1 = \frac{1}{\sigma^2} \left. \frac{\drm \left( \sigma(z)^2 \right)} {\drm z} \right|_{z=z_1} > 0 \quad \textrm{and} \quad  \Sigma_2 = \frac{1}{\sigma^2} \left.\frac{\drm \left( \sigma(z)^2 \right)}{\drm z}\right|_{z=z_2}<0.
\label{eq:Sigma1Sigma2_2TP}
\end{equation}
The eigenvalue constraint, obtained by requiring that the two one-turning-point solutions match in region III (defined by $z-z_1\gg k^{-2/3}$ and $z_2-z\gg k^{-2/3}$), is
\begin{equation}
k\int_{z_1}^{z_2}\sqrt{-\calA(z)}\drm z\approx \left(n+\frac{1}{2}\right)\pi.
\label{eq:eigenvalues_condition_2TP}
\end{equation}
Figure~\ref{fig:2tpproblem} depicts the solution of a two-turning-point problem for Case~1 of Section~\ref{sec:TE}, using equations (\ref{eq:WKB_solutions_2TP_general}) with $C=1$, $k=8000$ and $n=70$, thus implying (from equation~(\ref{eq:eigenvalues_condition_2TP})) $z_1=0.520$ and $z_2=0.859$. 

The solutions obtained in Sections~\ref{subsec:max_in_interior} and \ref{subsec:max_at_bdry} can also be derived from the WKB analysis presented in this appendix. The two-turning-point solution~(\ref{eq:WKB_solutions_2TP_general})--(\ref{eq:eigenvalues_condition_2TP}) with $n=0$ constitutes an asymptotic solution for Section~\ref{subsec:max_in_interior} (with $|z_2-z_1|\sim k^{-1/2}$, implying no oscillations in the midlayer). The one-turning-point solution defined by expressions~(\ref{eq:WKB_gen_solutions_2regions}), (\ref{eq:solution_regII}), (\ref{eq:constants_234}), (\ref{eq:constants_52}) and (\ref{eq:eigenvalues_condition_sigma_increasing}) for $n=0$ corresponds to the case $\sigma(z-z_0)\sim z-z_0$ in Section~\ref{subsec:max_at_bdry}.  


\end{document}